\documentclass[aps,prl,reprint,twocolumn,amsmath,amssymb,citeautoscript]{revtex4-1}
\usepackage{graphicx}
\usepackage{dcolumn}
\usepackage{bm}
\usepackage{latexsym,epsfig}
\usepackage{graphicx}
\usepackage{verbatim}
\usepackage{comment}
\usepackage{amsmath}
\usepackage{amssymb}
\usepackage{stmaryrd}
\usepackage{color}
\usepackage{epstopdf}
\usepackage{grffile}
\usepackage{lipsum}
\usepackage{enumitem}
\usepackage{soul}
\DeclareGraphicsExtensions{.eps}

\newcommand{\beq}{\begin{equation}}
\newcommand{\eeq}{\end{equation}}
\newcommand{\bea}{\begin{eqnarray}}
\newcommand{\eea}{\end{eqnarray}}
\newcommand{\ben}{\begin{eqnarray*}}
\newcommand{\een}{\end{eqnarray*}}
\newcommand{\bfig}{\begin{figure}}
\newcommand{\efig}{\end{figure}}

\newcommand{\ua}{\uparrow}
\newcommand{\da}{\downarrow}

\usepackage{hyperref}
\hypersetup{
    colorlinks=true,      
    urlcolor=blue,
    citecolor=blue,
    linkcolor=blue
}

\begin{document}
\title{Signatures of non-trivial pairing in the quantum walk of two-component bosons}
\author{Mrinal Kanti Giri$^1$, Suman Mondal$^1$, B. P. Das$^{2,3}$ and Tapan Mishra$^{1,2,4}$}
\email{mishratapan@gmail.com}
\affiliation{$^1$Department of Physics, Indian Institute of Technology, Guwahati-781039, India}
\affiliation{$^2$Centre for Quantum Engineering Research and Education,
TCG Centres for Research and Education in Science and Technology, Sector V, Salt Lake, Kolkata 70091, India}
\affiliation{$^3$ Department of Physics, School of Science, Tokyo Institute of Technology,
2-1-2-1-H86, Ookayama, Meguro-ku, Tokyo 152-8550, Japan}
\affiliation{$^4$ School of Physical Sciences, National Institute of Science Education and Research, HBNI, Jatni 752050, India}


\date{\today}

\begin{abstract}
Nearest neighbour bosons possessing only onsite interactions do not form onsite bound pairs in their quantum walk due to fermionization. We obtain signatures of non-trivial onsite pairing in the quantum walk of strongly interacting two component bosons in a one dimensional lattice. By considering an initial state with particles from different components located at the nearest-neighbour sites in the central region of the lattice, we show that in the dynamical evolution of the system, competing intra- and inter-component onsite repulsion leads to the formation of onsite inter-component bound states. We find that when the total number of particles is three,  an inter-component pair is favoured in the limit of equal intra- and inter-component interaction strengths. However, when two bosons from each species are considered, inter-component pairs and trimer are favoured depending on the ratios of the intra- and inter-component interactions. In both the cases, we find that the quantum walks exhibit a re-entrant behaviour as a function of inter-component interaction.

\end{abstract}





\maketitle
{\em Introduction.-}The phenomenon of Quantum Walk (QW) has attracted much attention recently due to its importance in fundamental science as well as quantum information~\cite{Kempe2003, Childs2013}. The propagation of quantum particles in different sites obeying the superposition principle of quantum mechanics makes the QW superior compared to its classical counterpart ~\cite{Aharonov}. Although there are two types of QW known as discrete and continuous-time QW that have been proposed~\cite{Kempe2003}, the framework of continuous-time QW (CTQW) provides a versatile approach to study the dynamical properties of quantum mechanical particles in a lattice at a few particle levels. Owing to their accessibility in both theoretical and experimental approaches, the CTQW has been studied and observed in disparate systems such as ion traps, trapped neutral atoms, photonic lattices, optical waveguides, and superconducting circuits~\cite{Schmitz2009,Zahringer2010,Karski2009,Weitenberg2011,Fukuhara2013,Manouchehri2014,Hoyer2010,Mohseni2008,Peruzzo2010,Lahini_walk,poulios2014quantum,Yan}. 

Recent studies on periodic lattices show that the QWs of more than one indistinguishable particle exhibits non-trivial correlations due to Hanbury Brown-Twiss (HBT) interference~\cite{Greiner_walk,Lahini_walk}. From theoretical and experimental studies, it is well established that when the walkers start from the same site, the individual particle wavepackets spread ballistically and symmetrically from their initial positions. However, when the walkers are at two nearest neighbour (NN) sites, they propagate together, exhibiting the phenomenon of bosonic bunching. Further developments in studying the QWs of interacting particles have enabled us to gain insights into the combined effects of interactions, particle statistics and strong correlations~\cite{etai_greiner,mondalwalk, sowinskiprb2020, PhysRevLett.125.165301, Roos2017}. Interestingly, the presence of interactions between the particles leads to a completely different scenario in the QW which has recently been studied in the context of the Bose-Hubbard models in one-dimension~\cite{Zakrzewski2017,Lahini_walk}. It has been shown that two strongly interacting bosons on a single site exhibit QW of bound bosonic pairs, whereas two NN bosons show a transition from bosonic to fermionic like spatial correlations and anti-bunching with an increase in onsite interaction~\cite{Greiner_walk,Yan}.
Moreover, the QW of two NN bosons in the presence of NN interaction exhibits the signatures of a NN pair~\cite{PhysRevA.90.062301, Luis-Arya}.
On the other hand, the QWs of two interacting distinguishable particles have also been explored in one dimension~\cite{Dias2007,Sowinski_sarkar,giri2021quantum,Dias2016} exhibiting features qualitatively similar to the QW of indistinguishable particles. 

An important inference that can be drawn from the existing findings is that strongly interacting NN bosons don't form local pairs in their QW. However, in this paper, we show that in the case of the QW of two-component bosons in one-dimension, non-trivial local bound pairs can be formed due to the interplay of inter-
and intra-component interactions.
By considering different initial states of more than two particles, we show that the quantum correlation along with competing interactions favors the formation of onsite inter-component bound pairs even if the two components start their QW from the NN sites.
Depending on the initial conditions the formation of bound pairs are found to be more robust when suitable hopping asymmetry is introduced. Moreover, we obtain a re-entrant feature in the QW as a function of the inter-species interaction.

{\em Model.-} Our analysis is based on the two-component Bose-Hubbard model which is given as 
\begin{eqnarray}
H=&-&\sum_{\langle i,j\rangle,\sigma}J_\sigma(a_{i,\sigma}^{\dagger} a_{j,\sigma}+H.c.)+
U_{\ua\da}\sum_{i}n_{i,\ua}n_{i,\da}\nonumber\\
&+&\sum_{i,\sigma} \frac{U_{\sigma}}{2} n_{i,\sigma}(n_{i,\sigma}-1)
\label{eq:ham}
\end{eqnarray} 
where $a_{i,\sigma}^{\dagger}$($a_{i,\sigma}$) are the creation (annihilation) operators of two different components $\sigma\in(\ua,\da)$ which can correspond to two different atoms or two hyperfine states of a single atom. 
$n_{i,\sigma}=a_{i,\sigma}^{\dagger} a_{i,\sigma}$ is 
the number operator at the $i^{\rm{th}}$ site corresponding to individual component $\sigma$. 
Here, $J_\sigma$ and $U_\sigma$ are the NN hopping matrix elements and onsite intra-component interaction energies of the individual components $\sigma$, respectively. The inter-component interaction is denoted by $U_{\ua\da}$.  In our studies, we assume the two components as the two hyperfine states of a single atom in a state-dependent optical lattice~\cite{Altman2003}. This assumption leads to the condition $U_\ua=U_\da = U$ and we define $\delta=J_\da/J_\ua$ to introduce the hopping asymmetry between the states/components. Hopping asymmetry is ensured by setting $J_\ua > J_\da$ i.e. $\delta < 1$ and we take  $J_\ua=1$ as the energy scale.

We study the CTQW (hereafter referred as QW) by computing experimentally relevant quantities such as the onsite densities and the two-site correlation functions. The total onsite density is defined as  
 $n_{i}(t)=\langle \sum_\sigma a_{i,\sigma}^{\dagger}a_{i,\sigma}\rangle$.
Unlike the single particle case~\cite{Lahini_walk,Greiner_walk,mondalwalk}, for the two-component system we compute both inter-species density-density and intra-species two-particle correlation function defined as 
\begin{equation}
\Gamma_{ij}^{\da\ua}(t) = \langle n_{i,\da} n_{j,\ua}\rangle ~~~ {and}~~~ 
\Gamma_{ij}^{\sigma}(t) = \langle a_{i,\sigma}^\dagger a_{j,\sigma}^\dagger a_{j,\sigma} a_{i,\sigma}\rangle
 \label{eq:gamma}
\end{equation}
respectively. These quantities are calculated with a time evolved state $|\Psi(t)\rangle=e^{-iHt/\hbar}|\Psi(0)\rangle$, where $|\Psi(0)\rangle$ is some initial state. 
The time evolution is obtained by utilizing the Time Evolving Block Decimation (TEBD) method using the Matrix Product States (MPS)~\cite{PhysRevLett.91.147902,PhysRevLett.93.040502} with appropriate  numerical control parameteres, given in ~\cite{TEBDapprox}. 
The simulations are carried out using the open source MPS (OSMPS) library~\cite{wall2015out,Jaschke_2018}. In our analysis, we consider a system size of $L=41$ except for the case of long time evolution where we take $L=82$.

\begin{figure}[b]
	\begin{center}
		\includegraphics[width=1.0\columnwidth]{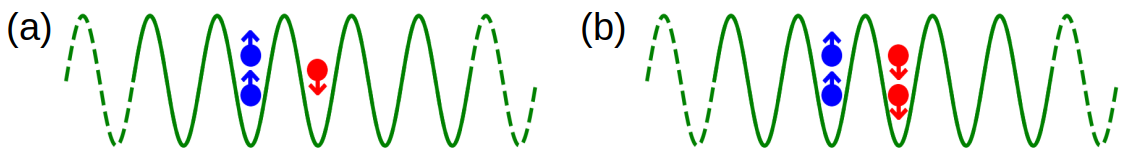}
	\end{center}
	\caption{The lattice diagrams (a) and (b) represent the initial state $|\Psi(0)\rangle_{\rm_{I}} = a_{0,\ua}^{\dagger 2} a_{1,\da}^{\dagger}|vac\rangle$ and $|\Psi(0)\rangle_{\rm_{II}} = a_{0,\ua}^{\dagger 2} a_{1,\da}^{\dagger 2}|vac\rangle$, respectively. }
	\label{fig:latt}
\end{figure}

For our studies we consider two different initial states where one or two particles from different components are located at the two NN sites at the center of the lattice. The states are (i) two $\ua$ particles and one $\da$ particle i.e. $|\Psi(0)\rangle_{\rm_{I}} = a_{0,\ua}^{\dagger 2} a_{1,\da}^{\dagger}|vac\rangle$ and (ii) two $\ua$ and two $\da$ particles i.e. $|\Psi(0)\rangle_{\rm_{II}} = a_{0,\ua}^{\dagger 2} a_{1,\da}^{\dagger 2}|vac\rangle$ as depicted in Fig.~\ref{fig:latt}(a) and (b) respectively.
In the following we discuss the QWs for both the cases in details.

\begin{figure}[t]
\centering
{\includegraphics[width=3.45in]{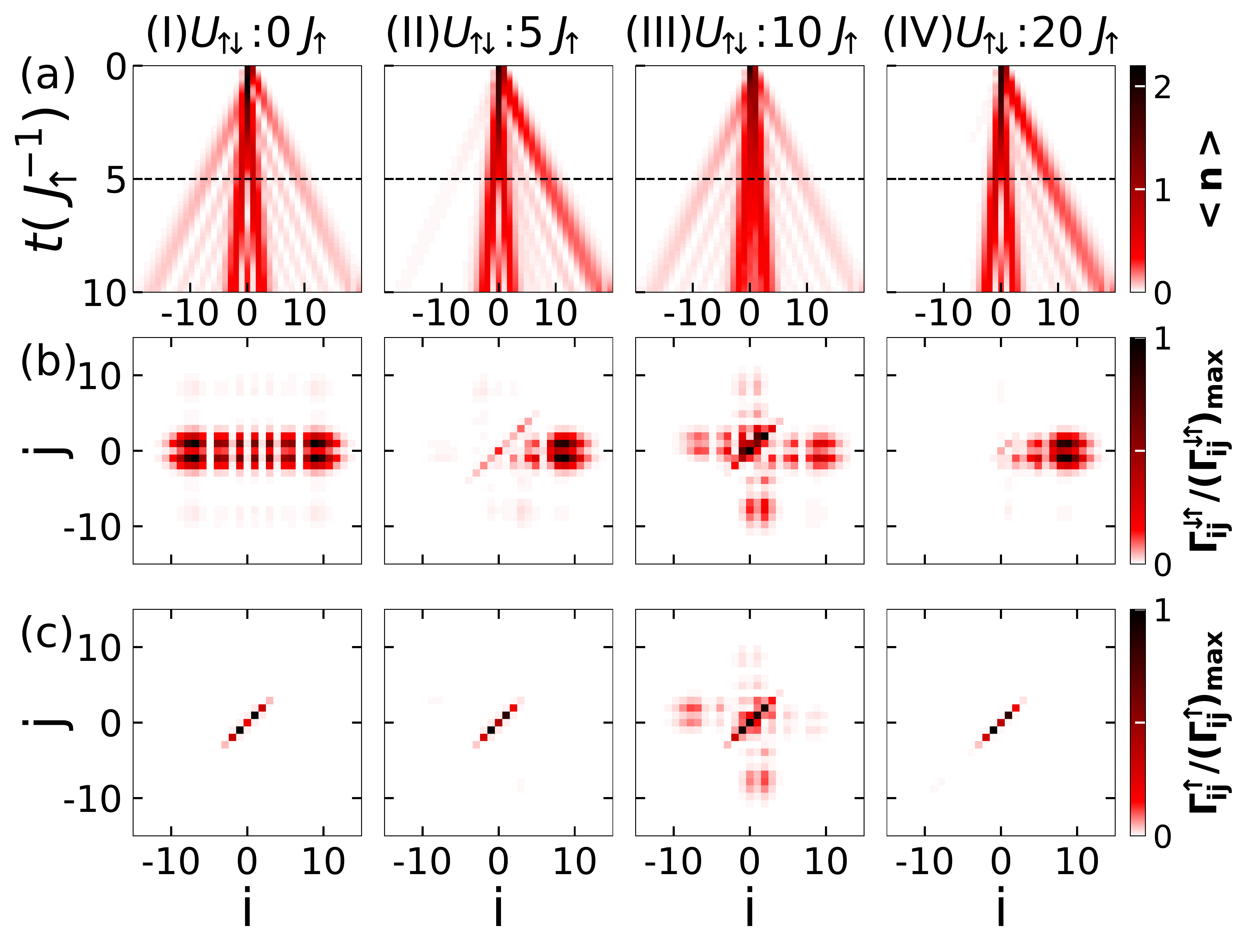}}
\caption{Panel (a) shows the onsite density evolution with the initial state $|\Psi(0)\rangle_{\rm {I}}$ for (I) $U_{\ua\da} = 0J_\ua$, (II) $U_{\ua\da}=5J_\ua$ , (III) $U_{\ua\da}=10J_\ua$ and (IV) $U_{\ua\da}=20J_\ua$. Panel (b) and (c) show the correlation functions $\Gamma_{ij}^{\da\ua}$ and $\Gamma_{ij}^{\ua}$, respectively. Here $U=10J_\ua$ and $\delta = 1$ are considered and the correlation functions are plotted at $t=5J_\ua^{-1}$.}
\label{fig:Fig1}
\end{figure}

{\em Two $\ua$ and one $\da$ particles.- }
In this case we consider two $\ua$ particles located at the central site (i.e. $i=0$) of the lattice and a $\da$ particle at the NN site on the right (i.e. $i=1$). The initial state corresponding to this situation is $|\Psi(0)\rangle_{\rm {I}}$ which is defined earlier.  This choice of the initial state ensures that $U_\da$ is irrelevant in the Hamiltonian of Eq.~\ref{eq:ham}. In such a scenario, the competing interactions are $U_\ua=U$ and $U_{\ua\da}$. We first discuss the symmetric hopping case i.e. $\delta=1$. In the absence of $U_{\ua\da}$, the two components behave independently in their QWs. For large $U$, the two $\ua$ particles form a repulsively bound pair~\cite{Winkler2006} and exhibit the QW of a composite particle with reduced hopping strength~\cite{Lahini_walk, Greiner_walk, Zakrzewski2017}. This situation is similar to the case of the QWs of two particles with asymmetric hopping as discussed in Ref.~\cite{giri2021quantum}. It is expected that with the onset of $U_{\ua\da}$, the individual wavepackets will start reflecting from each other leading to complete reflection in the limit of large $U_{\ua\da}$. In contrast, we show that for a moderate value of $U=10J_\ua$, which is sufficient to form a bound state of $\ua$ particles, the QW exhibits a re-entrant transition as a function of $U_{\ua\da}$ as can be seen from Fig.~\ref{fig:Fig1}(a)(I-IV). When $U_{\ua\da}=0J_\ua$, the QW shows a slow and fast spreading of densities indicative of that of $\ua\ua$ pair and $\da$ particle, respectively, which can be seen from the finite diagonal elements of the correlation matrix $\Gamma_{ij}^\ua$ as shown in Fig.~\ref{fig:Fig1}(c-I). An increase in $U_{\ua\da}$ leads to an onset of a fast spreading wavepacket reflected from the slower one (Fig.~\ref{fig:Fig1}(a-II)  for $U_{\ua\da}=5J_\ua$) a feature which reappears when $U_{\ua\da} > U$ (Fig.~\ref{fig:Fig1}(a-IV) for $U_{\ua\da}=20J_\ua$). This reflection of wavepacket is due to the inter-particle repulsion and can be understood from the vanishing of the upper triangular matrix elements of $\Gamma_{ij}^{\da\ua}$ as shown in Fig.~\ref{fig:Fig1}(b-II) and Fig.~\ref{fig:Fig1}(b-IV) plotted for $U_{\ua\da}=5J_\ua$ and $20J_\ua$ respectively. Careful analysis of the correlation function, however, reveals that while in the limit $U_{\ua\da}<U$ and $U_{\ua\da} > U$, the $\ua\ua$ pair survives (see Fig.~\ref{fig:Fig1}(c-II)) and Fig.~\ref{fig:Fig1}(c-IV), at $U_{\ua\da} \sim U$  it tends to break and a two-component pair (which we call a doublon i.e. $\ua\da$) tends to form - a scenario completely different from $U_{\ua\da}=0J_\ua$ limit (see Fig.~\ref{fig:Fig1}). This feature can be clearly seen from the gradual fading away of the diagonal elements of the intra-component correlation matrix $\Gamma_{ij}^\ua$ (Fig.~\ref{fig:Fig1}(c-III)) and appearance of finite diagonal elements of inter-component density correlation matrix $\Gamma_{ij}^{\da\ua}$ (Fig.~\ref{fig:Fig1}(b-III)). 

\begin{figure}[t]
\centering
{\includegraphics[width=1.0\columnwidth]{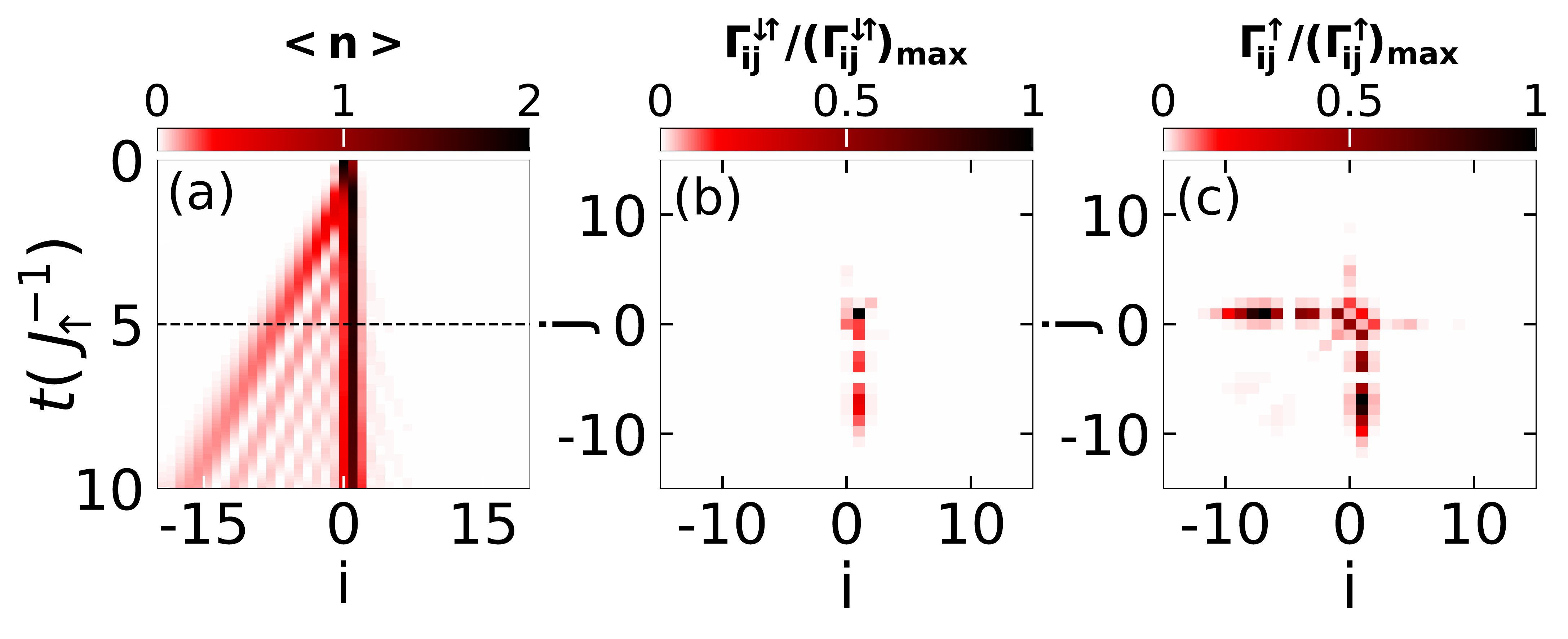}}
\caption{Figure shows (a) $\langle n_{i} \rangle$ , (b) $\Gamma_{ij}^{\da\ua}$ and (c) $\Gamma_{ij}^{\ua}$ for the QWs with the initial state $|\Psi(0)\rangle_{\rm {I}}$ for $U_{\ua\da} = 10J_\ua$, $U=10J_\ua$ and $\delta=0.2$. The correlation functions are plotted at $t=5J_\ua^{-1}$.}
\label{fig:Fig2}
\end{figure}

Interestingly, we further obtain that by reducing the hopping strength of the $\da$ component compared to the $\ua$ component, i.e., making $\delta < 1$, breaks the $\ua\ua$ pair completely and a stable doublon is formed after a short time evolution. This doublon acts as a potential barrier that reflects the wavepacket of the isolated $\ua$ component. These features can be seen from Fig.~\ref{fig:Fig2} where the density evolution and correlation functions are plotted by considering $\delta=0.2$ while keeping $U_{\ua\da}=U=10J_\ua$. Note that in this case also the re-entrant feature appears in the QW similar to the case of $\delta=1$ except a noticeable change in the spreading of densities due to reflection~\cite{supmat}.

In order to quantify the doublon formation and the dissociation of $\ua\ua$ pair, we compute the quantities defined as
\begin{equation}
 P_{\ua\da} = \sum_{i}n_{i,\ua} n_{i,\da} ;~~~ P_{\ua\ua} = 1/2\sum_{i}(n_{i,\ua}^2 - n_{i,\ua}),
\end{equation}
which count the number of $\ua\da$ and $\ua\ua$ pairs in the system and can be computed from the correlation matrix.

\begin{figure}[t]
\centering
{\includegraphics[width=3.45in]{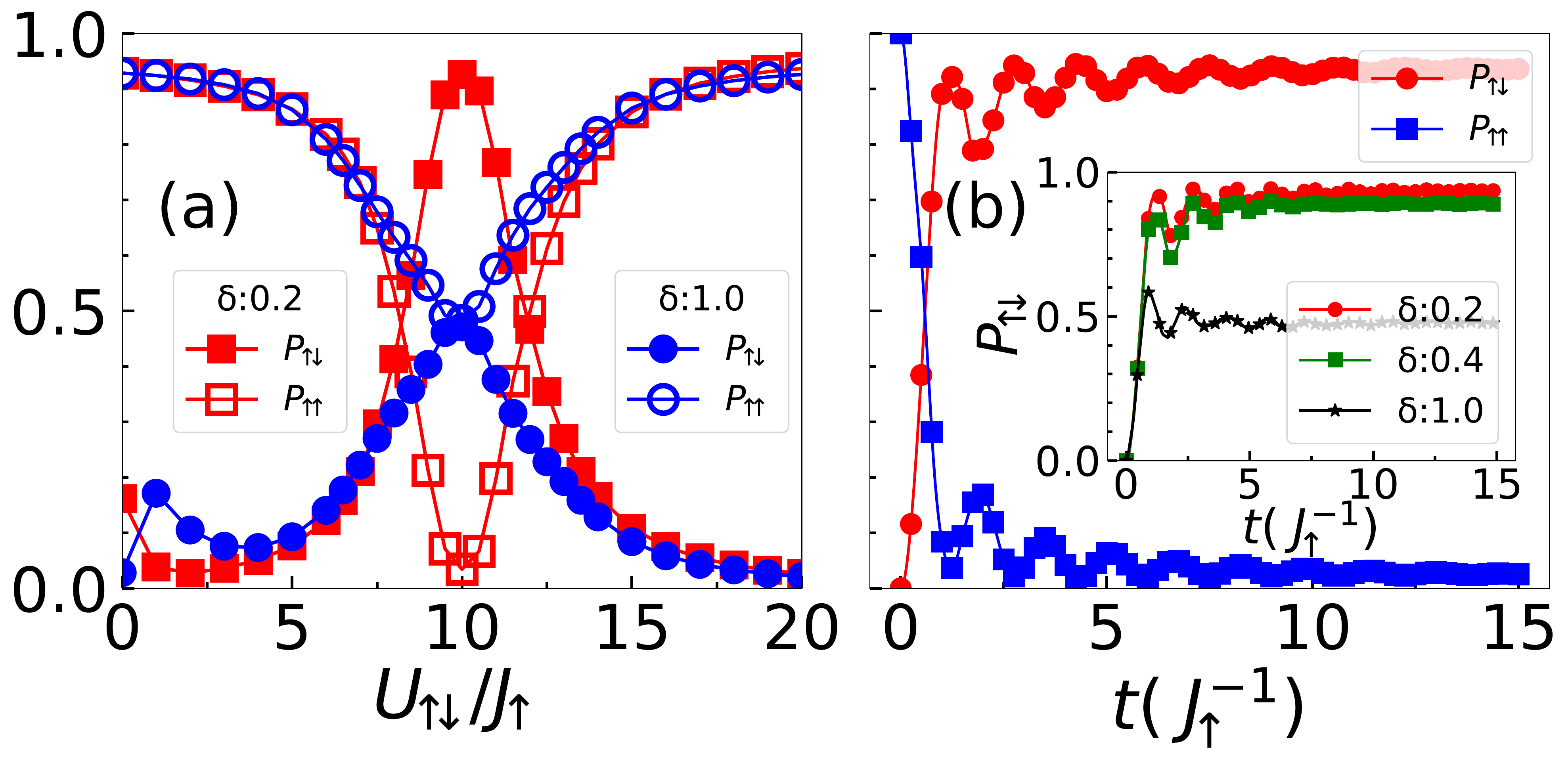}}
\caption{(a) $P_{\ua\da}$ (filled symbols) and $P_{\ua\ua}$ (open symbols) are plotted against $U_{\ua\da}/J_\ua$ for $\delta=0.2$ (red squares) and $\delta=1.0$ (blue circles) at $t=10J_\ua^{-1}$. (b) Shows the time evolution of $P_{\ua\da}$ (red circles) and $P_{\ua\ua}$ (blue squares) for $\delta=0.2$ and $U_{\ua\da}=10J_\ua$ indicating complete inter-component pair formation and breaking up of $\ua\ua$ pair. (Inset) The time evolution of $P_{\ua\da}$ for different $\delta$ such as $\delta=0.2$ (red circles), $\delta=0.4$ (green squares) and $\delta=1$ (black stars).}
\label{fig:Fig3}
\end{figure}

In  Fig.~\ref{fig:Fig3}(a), we plot both $P_{\ua\da}$ (filled symbols)  and $P_{\ua\ua}$ (open symbols) as a function of $U_{\ua\da}$ for both $\delta=0.2$ (red squares) and $\delta=1.0$ (blue circles) while keeping $U=10J_\ua$, after a time evolution to $t=10J_\ua^{-1}$. Clearly, the doublon formation is indicated by a dominant value of $P_{\ua\da}$ at $U_{\ua\da}=U=10J_\ua$ for $\delta=0.2$. Note that for $\delta=1$, both $P_{\ua\da}$ and $P_{\ua\ua}$ are of the same order due to the equal probabilities of formation of both the types of bound pairs. We also plot the time evolution of $P_{\ua\da}$ and $P_{\ua\ua}$ at the critical value $U_{\ua\da}=U=10J_\ua$ in Fig.~\ref{fig:Fig3}(b). The finite (zero) value of $P_{\ua\da}$ ($P_{\ua\ua}$) after $t > 1J_\ua^{-1}$ indicates the formation (dissociation) of $\ua\da$ ($\ua\ua$) pair. In the inset of Fig.~\ref{fig:Fig3}(b), the variation of $P_{\ua\da}$ for different values of $\delta$ confirms that the doublon formation is robust for smaller $\delta$. 

Summarizing up to this point, we have obtained that when $U_{\ua\da}$ is of the order of $U$, the $\ua\ua$ pair tends to break, and a $\ua\da$ pair tends to form. An introduction of hopping imbalance results in a complete dissociation of a $\ua\ua$ pair, and a doublon is formed. 

The reason behind this can be explained as follows. In the limit of equal inter and intra-species interaction and equal hopping strengths of both the components, the binding energy of $\ua\ua$ pair and $\ua\da$ pair are equal. Hence the states $|(\ua\ua)_0,~(\da)_1\rangle$ and $|(\ua)_0,~(\ua\da)_1\rangle$  are degenerate. Therefore, during the QWs, when the wavepacket of the $\ua\ua$ pair overlaps with that of the $\da$ component there is equal probability of forming either of the bound states. Hence, we see the signature of both $\ua\ua$ pair and $\ua\da$ pair in the QWs. However, by making the $J_\da$ smaller and comparable to the effective hopping strength of $\ua\ua$ pair, the doublon formation becomes energetically more favorable.  This is because the doublon formation increases the overall energy of the system, and the particles avoid each other due to repulsion. Note that this phenomenon is due to the interplay of both intra- and inter-species interactions and hence forbidden in the case of indistinguishable bosons~\cite{Zakrzewski2017} and two- component mixture with one particle from each species as considered in Ref.~\cite{giri2021quantum}.

\begin{figure}[t]
\centering
{\includegraphics[width=3.45in]{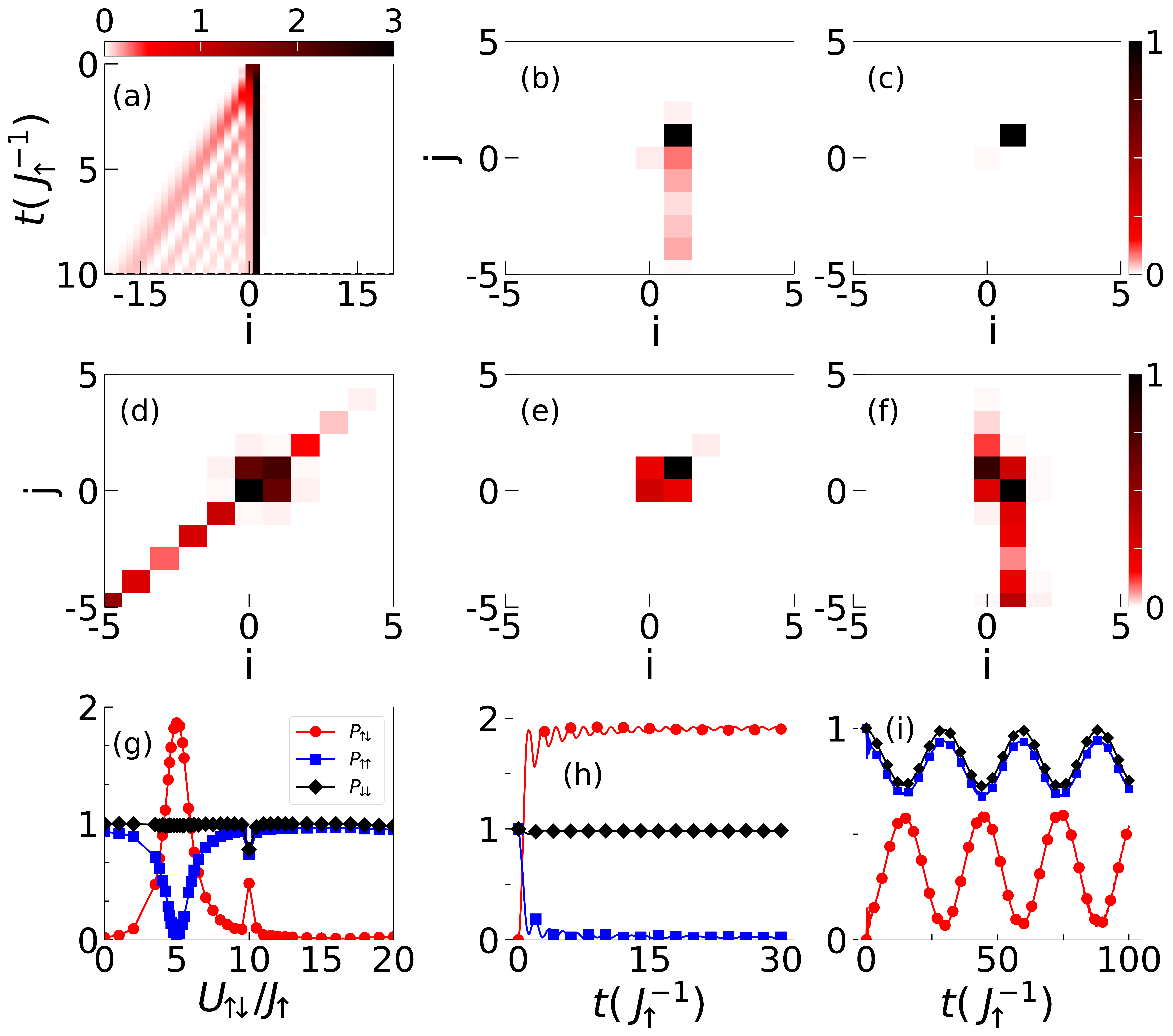}}
\caption{Figure shows (a)$\langle n_i \rangle$, (b)$\Gamma_{ij}^{\da\ua}$ and (c) $\Gamma_{ij}^{\da}$ for the QWs with the initial state $|\Psi(0)\rangle_{\rm {II}}$ for $U_{\ua\da} = 5J_\ua$, $U=10J_\ua$ and $\delta=0.2$ at $t=10J_\ua^{-1}$. (d-f) Show the values of $\Gamma_{ij}^{\ua}$, $\Gamma_{ij}^{\da}$ and $\Gamma_{ij}^{\da\ua}$, respectively for $U_{\ua\da}=10J_\ua$ at $t=17J_\ua^{-1}$. (g) Shows the behavior of $P_{\ua\da}$ (red circles), $P_{\ua\ua}$ (blue squares) and $P_{\da\da}$ (black diamond) as a function of $U_{\ua\da}$ at $t=10J_\ua^{-1}$. The time evolution of $P$'s are plotted in (h) and (i) for $U_{\ua\da}=5J_\ua$ and $10J_\ua$, respectively with $L=82$ sites.}
\label{fig:Fig4}
\end{figure}

{\em Two $\ua$ and two $\da$ particles.-}
In this part we consider two particles from each component and study  their QWs. The initial state considered for this case is given by $|\Psi_0\rangle_{\rm {II}} = a_{0,\ua}^{\dagger 2} a_{1,\da}^{\dagger 2}|vac\rangle$.
Note that for this initial state the $U_\da$ term in the Hamiltonian of Eq.~\ref{eq:ham} is
relevant which was ignored previously. Now, the physics of the system will be governed by all the three interactions, namely $U_\ua$, $U_\da$ and $U_{\ua\da}$. Similar to the previous cases, here we assume $U_\ua=U_\da=U=10J_\ua$ and vary $U_{\ua\da}$ for our investigation. We begin the discussion with asymmetric hopping and come back to the symmetric case later. In the limit of $U=10J_\ua$, both $\ua$ and $\da$ particles form repulsively bound pairs at the beginning when $U_{\ua\da}=0J_\ua$~\cite{Winkler2006,Greiner_walk,mondalwalk,Zakrzewski2017,Lahini_walk}. Upon increasing the value of $U_{\ua\da}$ with $\delta=0.2$,  we see simultaneous signatures of a three-particle and a single particle QW in the density evolution at $U_{\ua\da}=5J_\ua$ as shown in Fig.~\ref{fig:Fig4}(a). The figure indicates that a $\ua$ particle forms pair with an already formed $\da\da$ pair leaving behind an isolated $\ua$ particle indicated by the central bright patch. These features can be seen in the correlation data shown in Fig.~\ref{fig:Fig4}(b) and (c) where $\Gamma_{ij}^{\da\ua}$ and $\Gamma_{ij}^{\da}$ are plotted respectively. The bright spots at $(1,1)$ position in Fig.~\ref{fig:Fig4}(b) and (c) indicate the three-particle bound state. To further quantify this we compare the behaviour of $P_{\da\da} = 1/2\sum_{i} (n_{i,\da}^{2} - n_{i,\da})$ along with $P_{\ua\ua}$ and $P_{\ua\da}$ as a function of $U_{\ua\da}$ in Fig.~\ref{fig:Fig4}(g). The values of $P_{\ua\ua} \sim 0$ (blue square), $P_{\da\da} \sim 1$ (black diamond) and $P_{\ua\da}\sim 2$ (red circle) for $U_{\ua\da}=5J_\ua$ after the time evolution to $t=10J_\ua^{-1}$  confirm the formation of $\ua\da\da$ bound state, which we call a triplon. The triplon formation can also be confirmed from the time evolved values of $P$'s which saturate to  $P_{\ua\da}\sim 2$, $P_{\da\da} \sim 1$ and $P_{\ua\ua} \sim 0$ as shown in Fig.~\ref{fig:Fig4}(h). 
On the other hand the isolated $\ua$ particle can not penetrate the potential barrier created by the triplon and performs a unidirectional QW on the left part of the lattice as can be seen from Fig.~\ref{fig:Fig4}(a).

Further increase in $U_{\ua\da}$ tends to favour the formation of all possible pairs such as the doublon ($\ua\da$) and two intra-component pairs ($\ua\ua$ and $\da\da$) at $U_{\ua\da}=U=10J_\ua$. The signatures of which can be seen as the finite diagonal elements of the correlation matrices (Fig.~\ref{fig:Fig4}(d-f)) and the behaviour of the values of $P$'s (Fig.~\ref{fig:Fig4}(g and i)). It is to be noted that the $P$'s exhibit finite oscillation in their time evolution for $\delta=0.2$ which saturate fast when $\delta$ is increased~\cite{supmat}. The slow evolution for $\delta=0.2$ is due to the weak effective hopping. Interestingly, we also see the signature of a nearest-neighbour $\ua\da$ pair in the density-density correlation matrix in Fig.~\ref{fig:Fig4}(f). This unusual pairing is inevitable due to the simultaneous formation of onsite pairs and doublons.

The triplon formation at $U_{\ua\da}=U/2=5J_\ua$ can also be attributed to the condition of minimum effective  interaction which can be understood as follows. In the atomic limit, the states $|(\ua\ua)_0~(\da\da)_1\rangle$, $|(\ua)_0~(\ua\da\da)_1\rangle$ and $|(\ua\ua\da)_0~(\da)_1\rangle$ are degenerate. With $\delta=0.2$, the $\ua\ua$ pair is weakly bound compared to the $\da\da$ pair because $U/J_\ua < U/J_\da$. This ensures faster spreading of the former compared to the latter. Hence, during the time evolution, when the  wavepacket of the $\ua\ua$ pair overlaps with that of the $\da\da$ pair and due to degeneracy, a stable triplon is formed.
Once again, at $U_{\ua\da}=U=10J_\ua$, the system exhibits another condition of degenerate states where the particles prefer to be in states such as $|(\ua\da)_0~(\ua\da)_1\rangle$, $|(\ua\ua)_0~(\da\da)_1\rangle$. In the regime when $U_{\ua\da} < U=5J_\ua$, $5J_\ua < U_{\ua\da} < 10J_\ua$ and $U_{\ua\da}> 10J_\ua$ the bosons favors to stay in the original configuration of $|(\ua\ua)_0~(\da\da)_1\rangle$ without forming an inter-component bound state. This is because in the limit $U_{\ua\da} \gtrless U$ the breaking of the intra-component pair is energetically not favorable. Note that the re-entrant feature with respect to $U_{\ua\da}$ is also present in the four particle QW~\cite{supmat}.

On the other hand for $\delta=1$, the situation is completely different. Due to symmetric hopping strengths, the $\ua\ua$ and $\da\da$ pairs tend to break simultaneously and a doublon tends to form at a critical $U_{\ua\da}=5J_\ua$~\cite{supmat}. Note that similar to the QW with the initial state $\left| \Psi(0)\right \rangle_{\rm_{I}}$, the signature of doublon formation is weak in the four particle case with symmetric hopping strengths.

{\em Conclusions.-} 
In summary, our findings suggest a route to achieving local bound states in the QWs of initially non-local bosons with only local interactions in the context of the two-component Bose-Hubbard model. We have shown that the non-trivial inter-component bound states can be formed at certain critical ratios of inter and intra-component interaction strengths. By considering three particles in total, an inter-component bound pair is formed when both intra- and inter-component interactions are of equal strength. However, when two particles from each component are considered, a stable triplon is formed when the inter-species interaction is half of the intra-species ones. Moreover, we have obtained that a finite hopping asymmetry between the components plays an important role in favouring a more stable inter-component bound pair. We have also shown that the QWs exhibit a re-entrant phenomenon as a function of the inter-component interaction. 

The many-body physics of two different types of particles or two-component systems has been a topic of great interest in its own right~\cite{frahm2005} in condensed matter physics. Compared to the system with identical particles, the two-component systems are a much richer platform enabling access to a larger parameter space due to the presence of both intra- and inter-component interactions. The present analysis opens up possibilities for further exploration in the context of the quantum walk of two component bosons such as the effects of NN interactions and disorder. Due to the rapid progress in the manipulation of ultracold binary atomic mixture in optical lattices~\cite{Taglieber,Fermi-Fermi,Ospelkaus2006,Tilman2006,Best,Catani,Gadway}, many physical phenomena involving two-component bosons, fermions and Bose-Fermi mixtures have been predicted~\cite{Altman2003,Isacsson2005,Duan2003,Orth2008,Wang2016,Mathey2007,mishraps,Singh2017, SantosSSH2018,Mondal2020,Ye2016,PhysRevA.100.053607} and observed~\cite{Fukuhara2013,SSHHexptLe2020,PhysRevLett.121.130402} in the framework of the Hubbard and the two-component Bose-Hubbard models. Therefore, our findings can in principle, be simulated in a system of two-component Bose mixture in optical lattices by controlling the inter- and intra-component interactions by the Feshbach resonance and the individual hopping strengths by the state-dependent optical lattice~\cite{Altman2003,Duan_altman2003,Mandel_et_al2003,Soltan-Panahi2011,Jian-wei-2017}.

\bibliography{references}

\end{document}


\title{Supplementary material for "Signatures of non-trivial pairing in the quantum walk of two-component bosons"}
\author{Mrinal Kanti Giri$^1$, Suman Mondal$^1$, B. P. Das$^{2,3}$ and Tapan Mishra$^{1,2,4}$}
	\affiliation{$^1$Department of Physics, Indian Institute of Technology, Guwahati-781039, India}
	\affiliation{$^2$Centre for Quantum Engineering Research and Education,
		TCG Centres for Research and Education in Science and Technology, Sector V, Salt Lake, Kolkata 70091, India}
	\affiliation{$^3$ Department of Physics, School of Science, Tokyo Institute of Technology,
		2-1-2-1-H86, Ookayama, Meguro-ku, Tokyo 152-8550, Japan}
	\affiliation{$^4$ School of Physical Sciences, National Institute of Science Education and Research, HBNI, Jatni 752050, India}
	\date{\today}
	
\begin{abstract}
In this Supplementary Material we discuss the QW of two $\ua$ and one $\da$ particles for $\delta=0.2$. We also discuss the QW for the initial state with one $\ua$ and two $\da$ particles placed at the nearest-neighbour sites with hopping asymmetry. The re-entrant feature with respect to the inter-component interaction strength ($U_{\ua\da}$) is discussed in the four-particle QW. We discuss the QW of three and four particles case for attractive $U_{\ua\da}$, where the non-trivial pairing and re-entrant features are not feasible. 
We provide an perturbative analysis for the effective hopping strength of the triplon discussed in the main text.
\end{abstract}
	
	\maketitle



	\section{Two $\ua$ and one $\da$ particles}\label{subsec:3part1}
	\begin{figure}[b]
	\begin{center}
		\includegraphics[width=1.0\columnwidth]{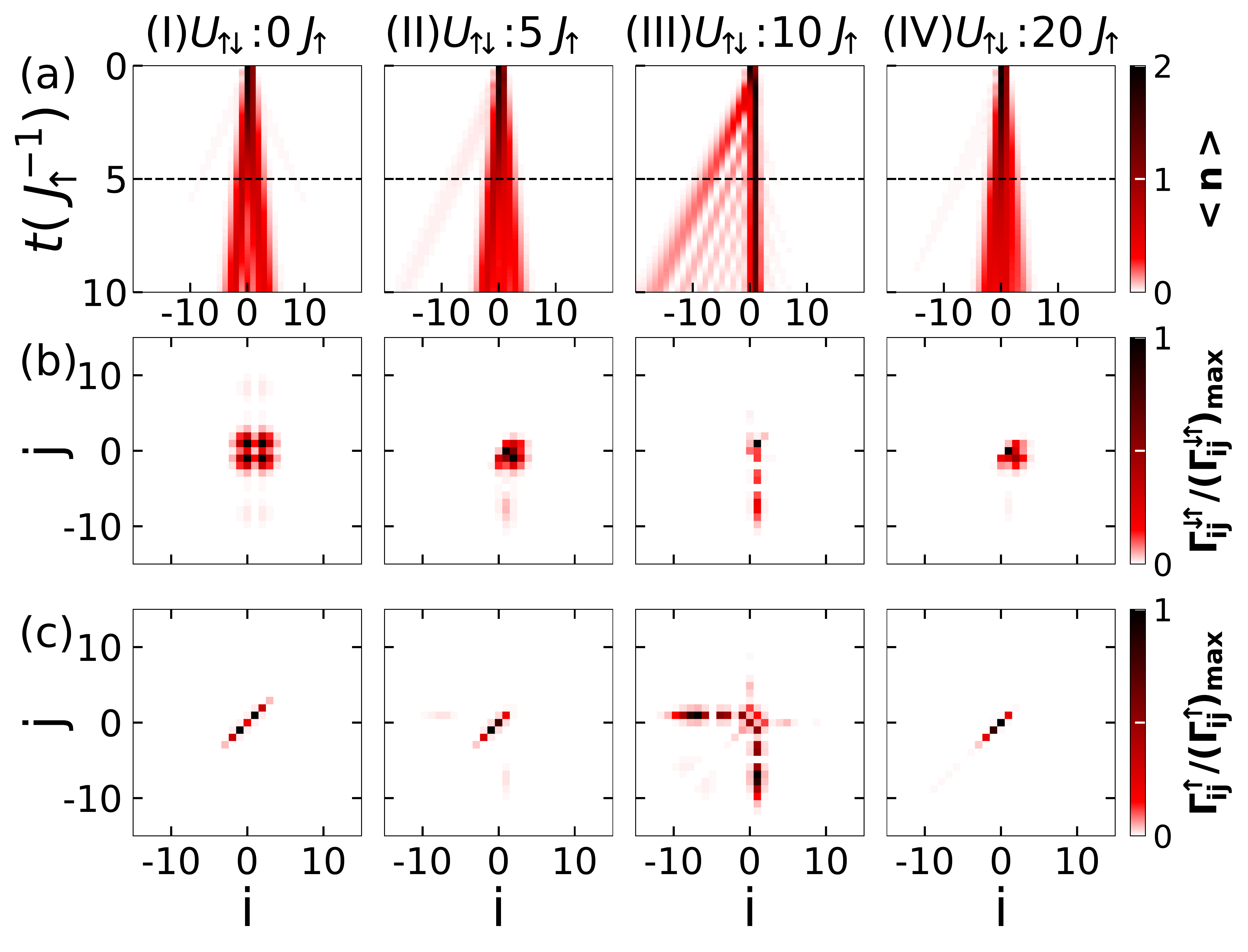}
	\end{center}
	\caption{Panel (a) shows the normalised density evolution of the initial state $|\Psi_0\rangle = a_{0,\ua}^{\dagger 2}a_{1,\da}^{\dagger}|vac\rangle$ for $U_{\ua\da} = 0J_\ua$, $5J_\ua$, $10J_\ua$ and $20J_\ua$. Here $U=10J_\ua$ and $\delta = 0.2$. Panel (b) and (c) are shown the correlation functions $\Gamma^{\da\ua}_{ij}$ and $\Gamma^{\ua}_{ij}$ respectively at time, $t=5J_\ua^{-1}$.}
	\label{fig:3den_cor}
	\label{fig:2l1hden_cor}
	\end{figure}

	\begin{figure}[b]
		\begin{center}
			\includegraphics[width=1.0\columnwidth]{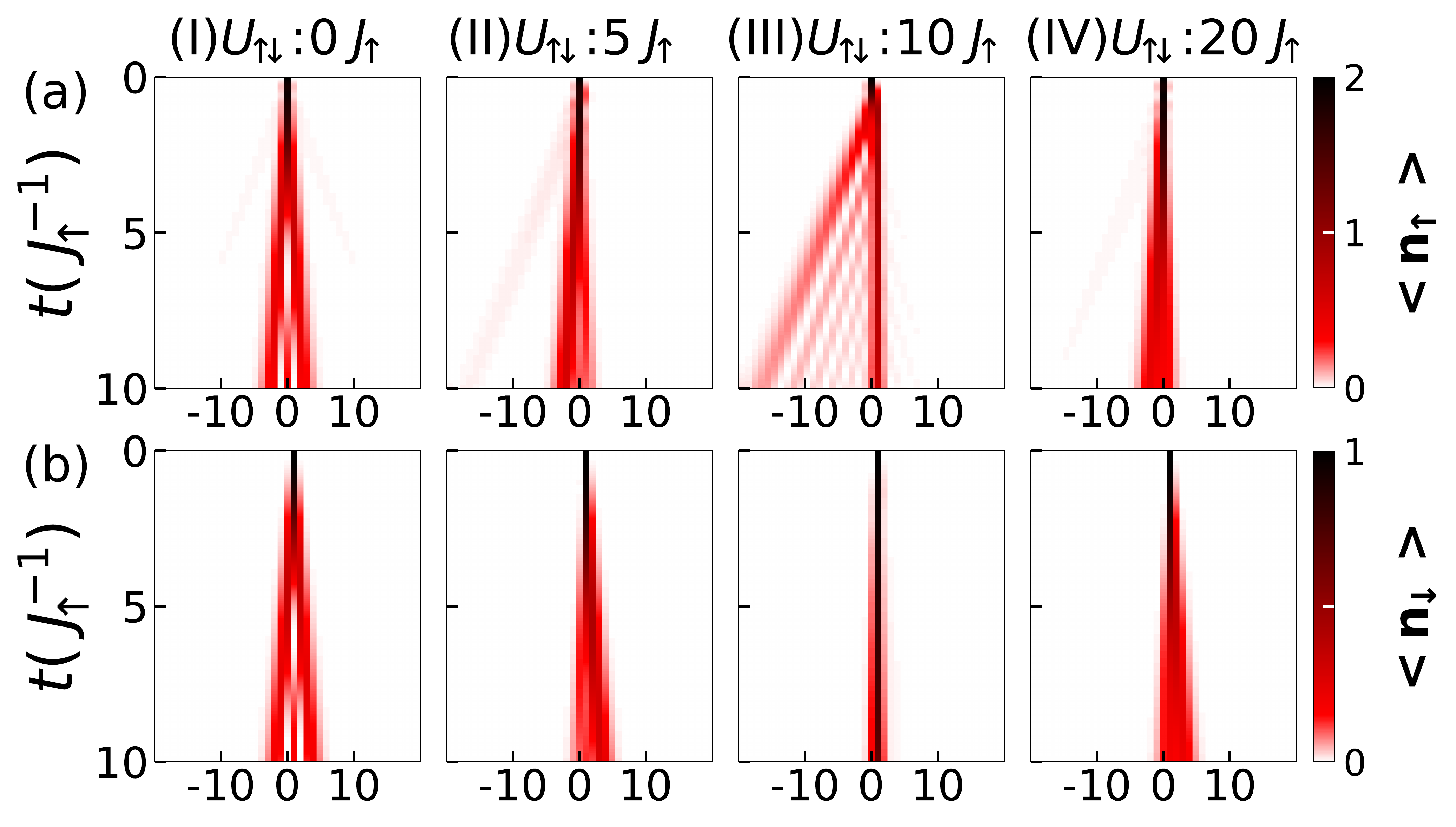}
		\end{center}
		\caption{Panel (a) and (b) shows the normalised density evolution of $n_{\ua}$ and $n_{\da}$, respectively for $U_{\ua\da} = 0J_\ua$, $5J_\ua$, $10J_\ua$ and $20J_\ua$. Here $U=10J_\ua$ and $\delta = 0.2$.}
		\label{fig:2l1hden}
	\end{figure}
	
	As discussed in the main text, the initial state for this scenario is $|\Psi(0)\rangle_{\rm_{I}} = a_{0,\ua}^{\dagger 2} a_{1,\da}^{\dagger}|vac\rangle$. 
	We plot the evolution of the onsite density ($\left\langle n_{i}(t)\right\rangle $), the density-density correlation ($\Gamma_{ij}^{\da\ua}$) and the two-particle correlation ($\Gamma_{ij}^{\ua}$) in Fig.~\ref{fig:2l1hden_cor} for $\delta=0.2$. 
	When $U_{\ua\da}=0J_\ua$, the QW shows a slow and fast spreading of densities (Fig.~\ref{fig:2l1hden_cor}(a-I)corresponding to the $\ua\ua$ pair and $\da$ particle respectively which can be seen from the finite diagonal elements of the correlation matrix $\Gamma^{\ua}_{ij}$ as shown in Fig.~\ref{fig:2l1hden_cor}(c-I). Increasing the value of $U_{\ua\da}$, the fast spreading wavepacket gets reflected from the slower one (Fig.~\ref{fig:2l1hden_cor}(a-II) for $U_{\ua\da}=5J_\ua$), a feature which reappears when $U_{\ua\da}>U$ (Fig.~\ref{fig:2l1hden_cor}(a-IV) for $U_{\ua\da}=20J_\ua$) indicating the re-entrant feature in the QW. The reflection of wavepacket in this case is due to the inter-particle repulsion and can be understood from vanishing of the upper triangular matrix elements of $\Gamma_{ij}^{\da\ua}$ as depicted in Fig.~\ref{fig:2l1hden_cor}(b-II) and Fig.~\ref{fig:2l1hden_cor}(b-IV) for $U_{\ua\da}=5J_\ua$ and $20J_\ua$ respectively. In the limit $U_{\ua\da}<U$ and $U_{\ua\da}>U$, the $\ua\ua$ pair survives (Fig.~\ref{fig:2l1hden_cor}(c-I), (c-II) and (c-IV)). Only for $U_{\ua\da} = U = 10J_\ua$, $\ua\ua$ pair tends to break and two component pair (doublon) tends to form (Fig.~\ref{fig:2l1hden_cor}(III)). For clarity we also show the density evolutions of $\ua$ and $\da$ components in Fig.~\ref{fig:2l1hden} for $\delta=0.2$. It can be seen that for $U_{\ua\da}=10J_\ua$, the $\ua\ua$ pair break into two different profiles. One of them remain close to the center which forms a doublon with the $\da$ component and at site $"1"$ and the other moves independently to the left (Fig.~\ref{fig:2l1hden}(a(III))).

	\begin{figure}[t]
		\begin{center}
			\includegraphics[width=1.0\columnwidth]{latt_2.png}
		\end{center}
		\caption{The lattice diagram represents the initial state $|\Psi(0)\rangle_{\rm_{III}} = a_{0,\ua}^{\dagger} a_{1,\da}^{\dagger 2}|vac\rangle$. }
		\label{fig:latt}
	\end{figure}

	\begin{figure}[b]
		\begin{center}
			\includegraphics[width=1.0\columnwidth]{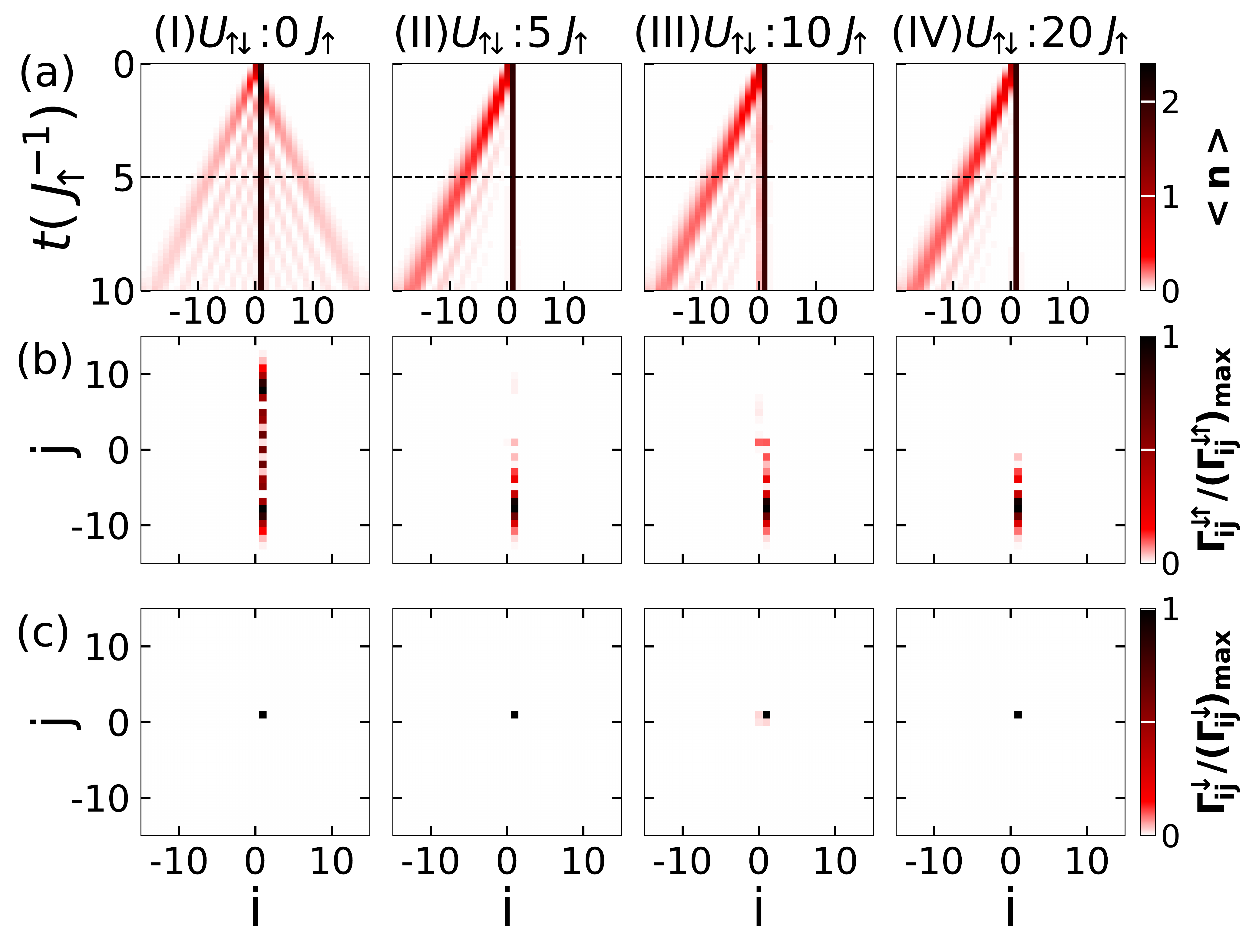}
		\end{center}
		\caption{Panel (a) shows the normalised density evolution of the initial state $|\Psi_0\rangle = a_{0,\ua}^{\dagger}a_{1,\da}^{\dagger 2}|vac\rangle$ for $U_{\ua\da} = 0J_\ua$, $5J_\ua$ $10J_\ua$ and $20J_\ua$. Here $U=10J_\ua$ and $\delta = 0.2$. Panel (b) and (c) are shown the correlation functions $\Gamma^{\da\ua}_{ij}$ and $\Gamma^{\da}_{ij}$ respectively at time, $t=5J_\ua^{-1}$.}
		\label{fig:2h1lden_cor}
	\end{figure}

	\begin{figure}[t]
		\begin{center}
			\includegraphics[width=1.0\columnwidth]{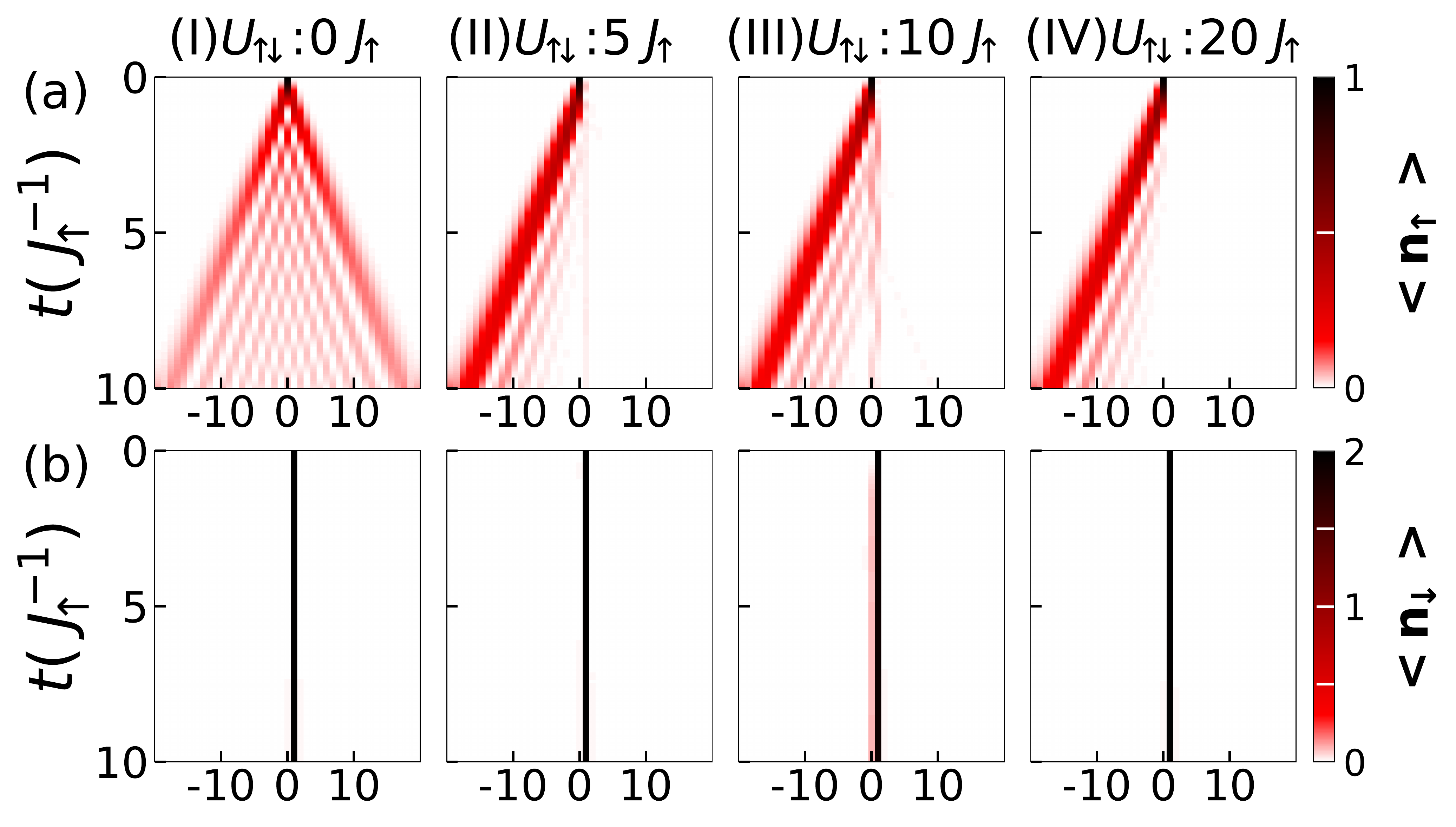}
		\end{center}
		\caption{Panel (a) and (b) shows the normalised density evolution of $n_\ua$ and $n_{\da}$, respectively for $U_{\ua\da} = 0$, $5$ $10$ and $20$. Here $U=10$ and $\delta = 0.2$.}
		\label{fig:2h1lden}
	\end{figure}

	\section{One $\ua$ and two $\da$ particles}\label{subsec:3part2}
	In this section we study the QW with a different initial state $|\Psi(0)\rangle_{\rm_{III}} = a_{0,\ua}^{\dagger} a_{1,\da}^{\dagger 2}|vac\rangle$  where we consider two $\da$ particle and one $\ua$ particle as shown in Fig.~\ref{fig:latt}. Although we have in total three-particles in the system, the situation is completely different from the QW with the initial state $|\Psi(0)\rangle_{\rm_{I}}$ when $\delta \neq 1$. In Fig.~\ref{fig:2h1lden_cor}, we plot the onsite density and the correlation functions for $U_\da=U=10J_\ua$ and $\delta=0.2$ for different values of $U_{\ua\da}$. It can be seen from Fig.~\ref{fig:2h1lden_cor}(a) that when $U_{\ua\da}=0J_\ua$, the two $\da$ particles exhibit slow spreading of densities corresponding to $\da\da$ pair localized around the 1$^{st}$ site and the single $\ua$ particle exhibits a fast propagation. Finite $U_{\ua\da}$ leads to the reflection of the $\ua$ component from the potential barrier created by the $\da\da$ pair. The signatures of the reflection of wavepacket and the $\da\da$ pair can be seen from the correlation matrix $\Gamma_{ij}^{\da\ua}$ and $\Gamma_{ij}^{\da}$ plotted in Fig.~\ref{fig:2h1lden_cor}(b) and (c) respectively. This phenomenon can also be seen by individually plotting the density evolution of the $\ua$ and $\da$ components as shown in Fig.~\ref{fig:2h1lden}.
	Note that the formation of a doublon ($\ua\da$ pair) in this case is forbidden due to the strong $\da\da$ pair.
	
	\begin{figure}[b]
		\begin{center}
			\includegraphics[width=1.0\columnwidth]{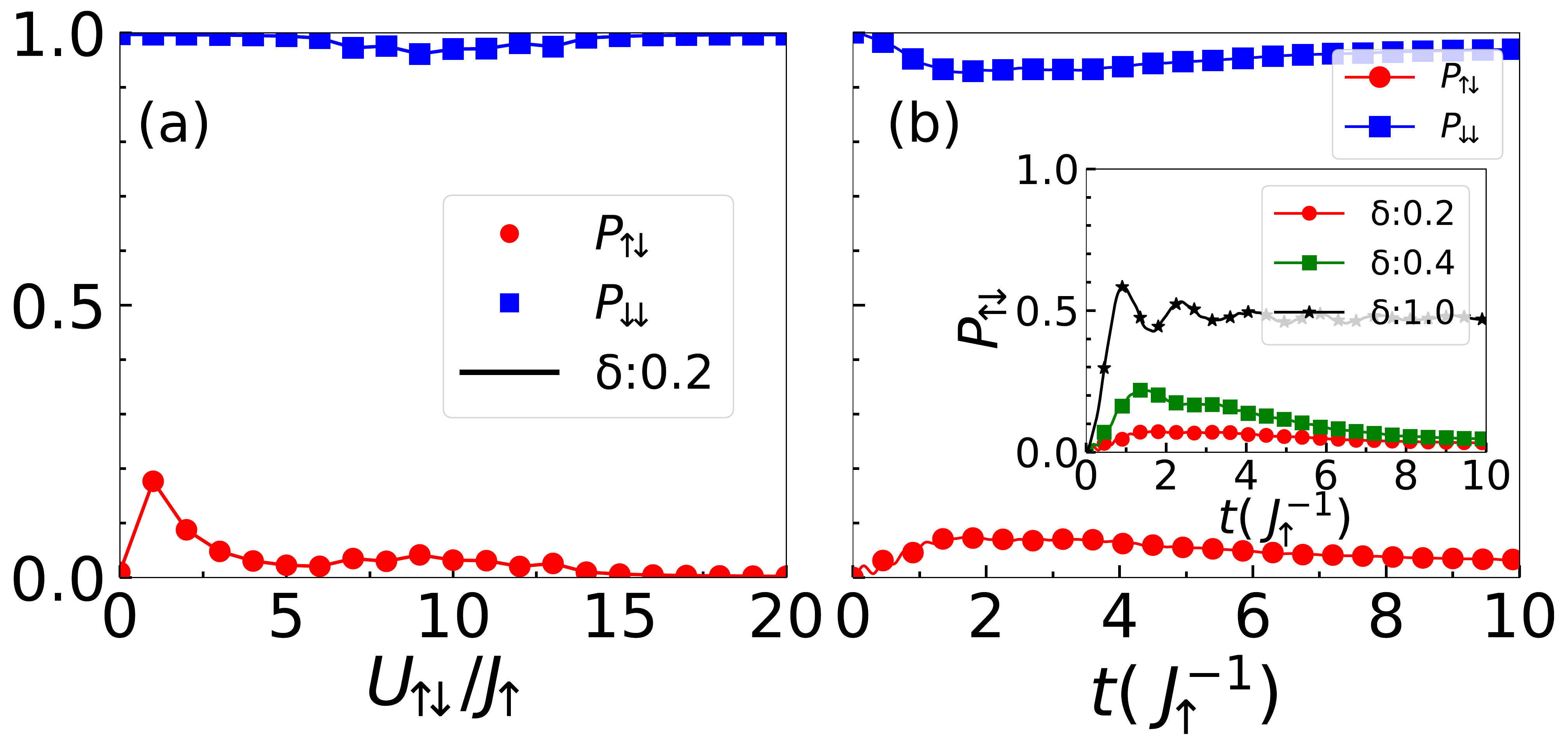}
		\end{center}
		\caption{(a)$P_{\ua\da}$(circles) and $P_{\da\da}$ (squares) against $U_{\ua\da}/J_\ua$ for $\delta=1$ (dashed curve) and $\delta=0.2$ (solid curve) at $t=10J_\ua^{-1}$. (b) Shows the time evolution of $P_{\ua\da}$ (circles) and $P_{\da\da}$ (squares) for different $\delta=0.2$ and $U_{\ua\da}=10J_\ua$ indicating that there are no inter-component pair formation and dissociation of $\da\da$ pair. (Inset) The time evolution of $P_{\ua\da}$ for different $\delta$ such as $\delta=0.2$ (circle), $\delta=0.4$ (square) and $\delta=1$ (star).}
		\label{fig:2h1lcor_vs_u}
	\end{figure}
	
	We also confirm the absence of doublon formation in the limit $\delta\neq1$ by comparing the $P$ values defined in the main text. In Fig.~\ref{fig:2h1lcor_vs_u}(a) we plot $P_{\ua\da}$ and $P_{\da\da}$ for different values of $U_{\ua\da}$ and $\delta=0.2$ after a time evolution to $t=10J_\ua^{-1}$. The values of $P_{\da\da}\sim 0$ and $P_{\ua\da}\sim 1$ as a function of $U_{\ua\da}$ indicate no doublon formation. This behaviour can be complemented by the time evolution of $P_{\da\da}$ and $P_{\ua\da}$ which saturate to $1$ and $0$ respectively (Fig.~\ref{fig:2h1lcor_vs_u}(b)). In the inset of Fig.~\ref{fig:2h1lcor_vs_u}(b) we show the time evolution of $P_{\ua\da}$ which saturates to a finite value for $\delta=1$ and vanishes as $\delta$ decreases. This indicates a behaviour opposite to that with the initial state $|\Psi(0)\rangle_{\rm_{I}}$ discussed in the main text.

	\begin{figure}[t]
		\centering
		{\includegraphics[width=0.5\textwidth]{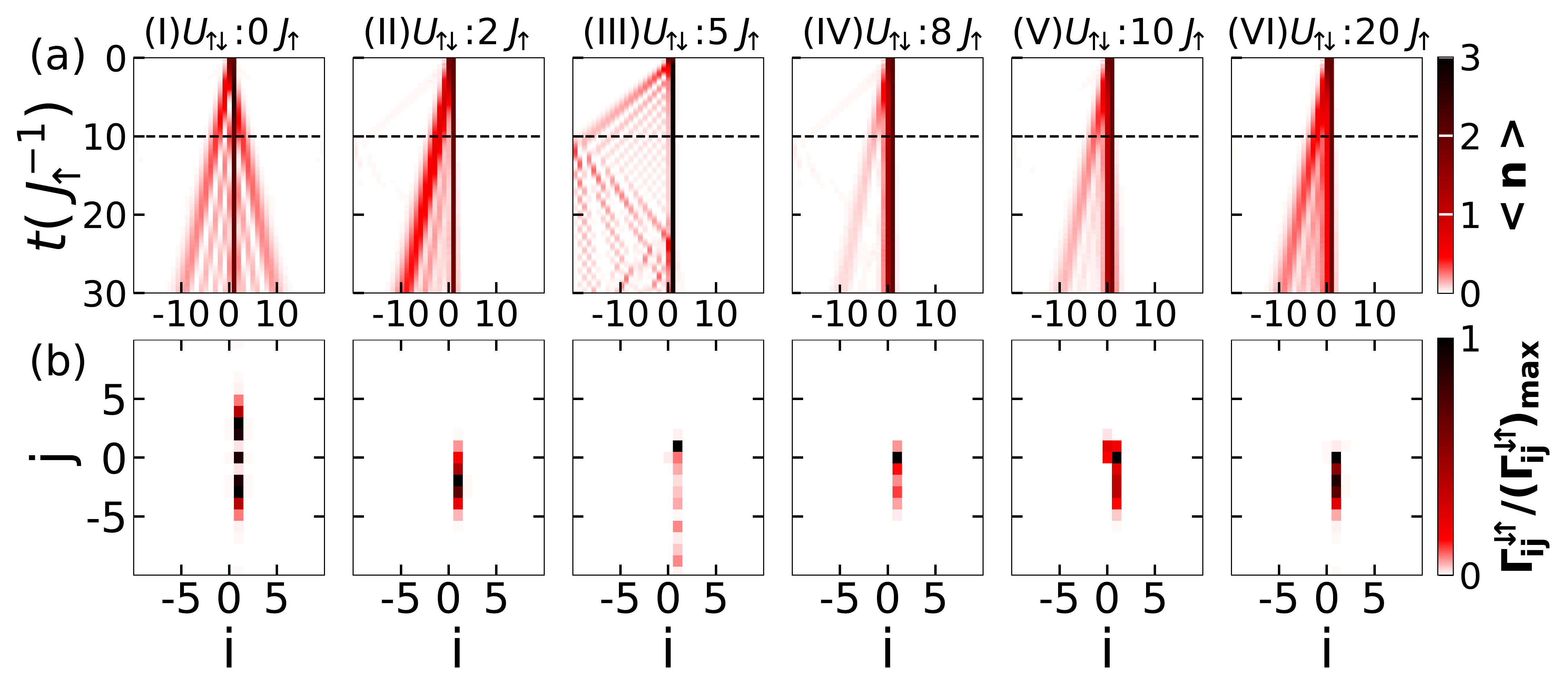}}
		\caption{Panel (a) shows the density evolution of the initial state $|\Psi_0\rangle = a_{0,\ua}^{\dagger 2} a_{1,\da}^{\dagger 2}|vac\rangle$ for $U_{\ua\da} = 0J_\ua, 2J_\ua, 5J_\ua, 8J_\ua$, $10J_\ua$ and $20J_\ua$. Here $U=10J_\ua$ and $\delta = 0.2$. Panel (b) shows the correlation functions $\Gamma^{\da\ua}_{ij}$ at time, $t=10J_\ua^{-1}$.}
		\label{fig:4Par_delta_0.2}
	\end{figure}
	
	\begin{figure}[b]
		\centering
		{\includegraphics[width=0.5\textwidth]{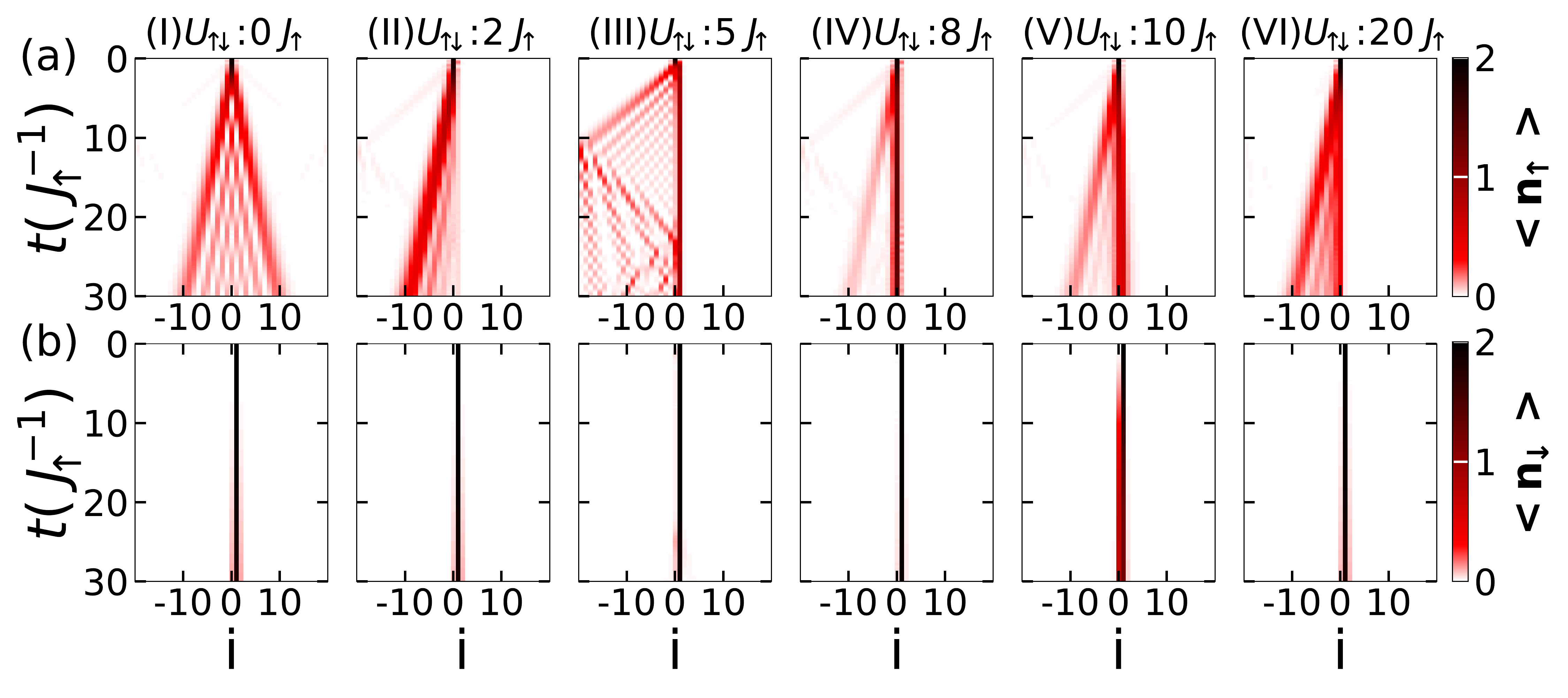}}
		\caption{Panel (a) and (b) shows the density evolution of $n_{\ua}$ and $n_{\da}$, respectively for $U_{\ua\da} = 0J_\ua, 2J_\ua, 5J_\ua, 8J_\ua$, $10J_\ua$ and $20J_\ua$. Here $U=10J_\ua$ and $\delta = 0.2$.}
		\label{fig:4Par_den}
	\end{figure}
	
	\begin{figure}[t]
		\centering
		{\includegraphics[width=0.5\textwidth]{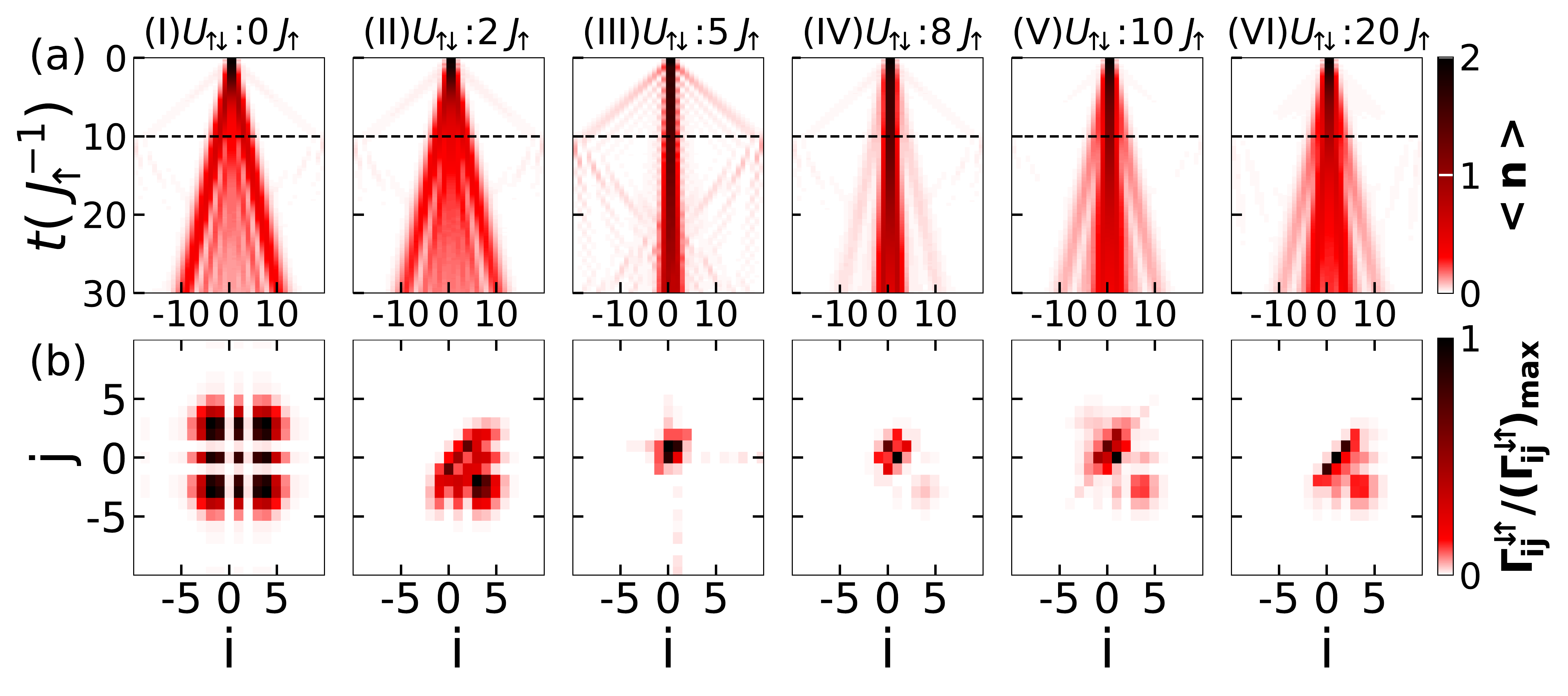}}
		\caption{Panel (a) shows the density evolution of the initial state $|\Psi_0\rangle = a_{0,\ua}^{\dagger 2} a_{1,\da}^{\dagger 2}|vac\rangle$ for $U_{\ua\da} = 0J_\ua, 2J_\ua, 5J_\ua, 8J_\ua$, $10J_\ua$ and $20J_\ua$. Here $U=10J_\ua$ and $\delta = 1.0$. Panel (b) shows the correlation functions $\Gamma^{\da\ua}_{ij}$ at time, $t=10J_\ua^{-1}$.}
		\label{fig:4Par_delta_1.0}
	\end{figure}
	
	\section{Two $\ua$ and two $\da$ particles}\label{subsec:4part}
	\subsection{Re-entrant features}
	In this section we show the features of re-entrant transition which appears in the QW of $ |\Psi_0\rangle_{\rm{II}} = a_{0,\ua}^{\dagger 2} a_{1,\da}^{\dagger 2}|vac\rangle$. We plot the density evolution ($\left\langle n_{i}(t)\right\rangle $), the density-density correlation function ($\Gamma_{ij}^{\da\ua}$) for $U=10J_\ua$ and $U_{\ua\da}=0J_\ua,~2J_\ua,~5J_\ua,~8J_\ua,~10J_\ua,~20J_\ua$ in Fig.~\ref{fig:4Par_delta_0.2} and Fig.~\ref{fig:4Par_delta_1.0} for $\delta=0.2$ and $\delta=1.0$ respectively. For $U_{\ua\da}=0J_\ua$ and $\delta=0.2$, the two $\ua\ua$ and $\da\da$ pairs perform QWs independently with fast and slow spreading respectively because of the difference in their effective tunneling rates. As $U_{\ua\da}$ increases, the fast moving wavepacket starts to get reflected from the slower one (Fig.~\ref{fig:4Par_delta_0.2}(a-II) for $U_{\ua\da}=2J_\ua$) which reappears in the limit when $U_{\ua\da}>U$ (Fig.~\ref{fig:4Par_delta_0.2}(a-VI) for $U_{\ua\da}=20J_\ua$). This reflection of wavepacket can be easily understood from the vanishing of the upper triangular matrix elements of $\Gamma^{\ua\da}_{ij}$ as shown in Fig.~\ref{fig:4Par_delta_0.2}(b-II) and Fig.~\ref{fig:4Par_delta_0.2}(b-VI). We plot individual density evolution $n_{\ua}$ and $n_{\da}$ in Fig.~\ref{fig:4Par_den} to observe the features i.e., reflection of the $\ua$ particle wavepacket, breaking of the $\ua\ua$ pair at $0\nth$ site and formation of pair at $1\st$ site.
	Similar features are also visible in the symmetric hopping case i.e. $\delta=1.0$ (see Fig.~\ref{fig:4Par_delta_1.0}(b-II) and (b-VI)). This indicates that the re-entrant feature in the QW of $\ua\ua$ and $\da\da$ pair appears in the limit of $U_{\ua\da} < 5J_\ua$ and $U_{\ua\da} > 10J_\ua$.
	
	\begin{figure}[!b]
		\begin{center}
			\includegraphics[width=0.95\columnwidth]{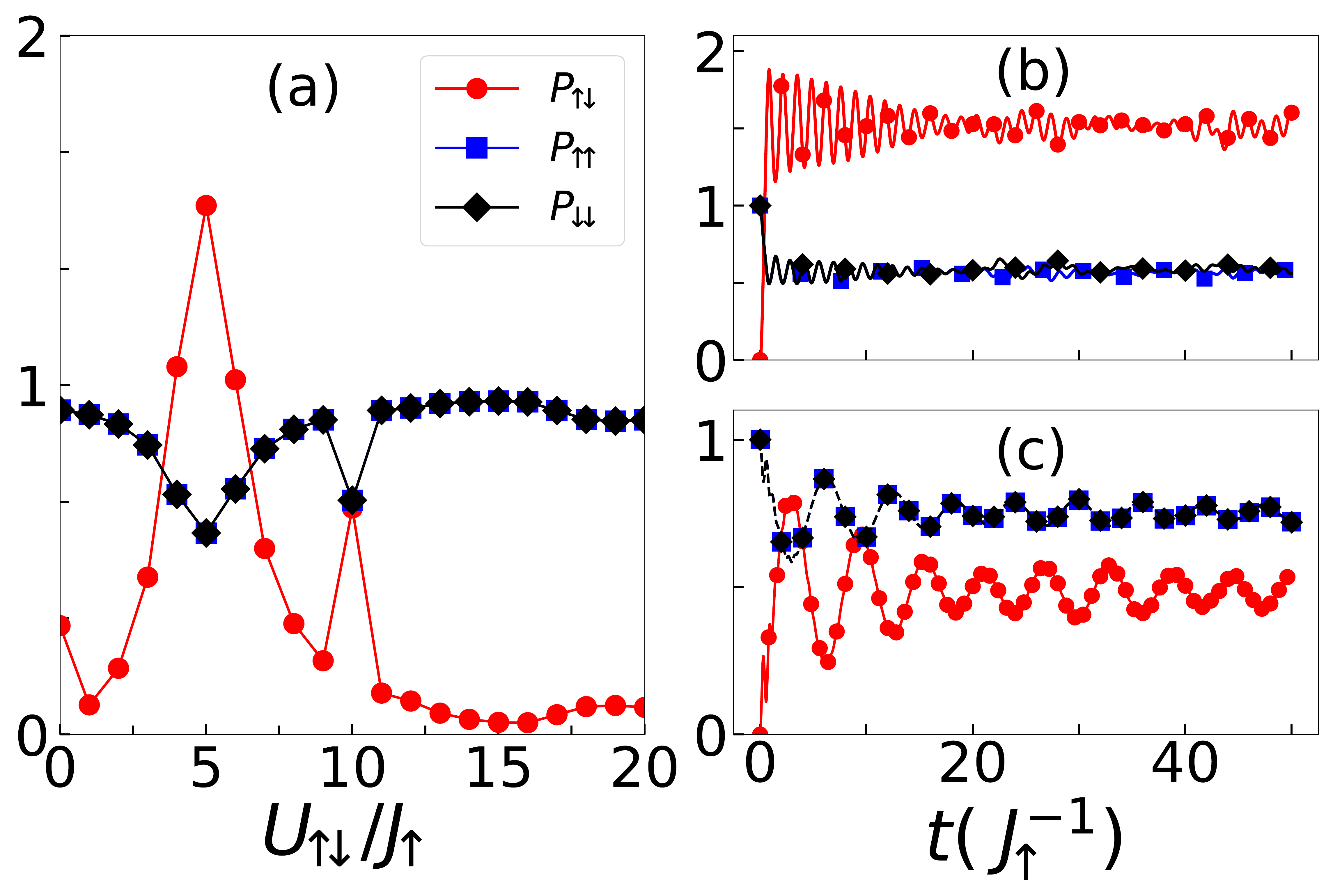}
		\end{center}
		\caption{(a) $P_{\ua\da}$ (red circles), $P_{\ua\ua}$ (blue squares) and $P_{\da\da}$ (black diamonds) against $U_{\ua\da}$ for $\delta=1$ at $t=10J_\ua^{-1}$. (b) Shows the time evolution of $P_{\ua\da}$ (red circles), $P_{\ua\ua}$ (blue squares) and $P_{\da\da}$ (black diamonds) for $U_{\ua\da}=5J_\ua$ indicating the formation of inter-component pair and dissociation of $\da\da$ pair. (c) The time evolution of $P_{\ua\da}$ (red circles), $P_{\ua\ua}$ (blue squares ) and $P_{\da\da}$ (black diamonds) for $U_{\ua\da}=10J_\ua$.}
		\label{fig:Pab_vs_time}
	\end{figure}
	
	\subsection{Pair formation}
	To find the difference between symmetric and asymmetric hopping cases, similar to Fig. 5(g) of the main text, we plot the $P$'s at time $t = 10 J_\ua^{-1}$  in Fig.~\ref{fig:Pab_vs_time}(a) for the $\delta = 1$. This indicates a similar behaviour compared to $\delta=0.2$ except at $U_{\ua\da} \sim 5J_\ua$. At $U_{\ua\da}=5J_\ua$, the values of $P_{\ua\da} < 2$ and $P_{\ua} = P_{\da} < 1$ indicate the partial formation (dissociation) of the triplon (pairs). Moreover, we plot  the variation of $P$'s with time $t(J_\ua^{-1})$ in Fig.~\ref{fig:Pab_vs_time}(b) which confirms the formation (dissociation) of triplon (pairs) with time. At $U_{\ua\da} \sim 10J_\ua$ on the other hand, the behaviour is same as the case of $\delta  =0.2$ (see Fig. 5 of the main text) besides the fact that the oscillation of $P$'s saturates faster to some values as shown in Fig.~\ref{fig:Pab_vs_time}(c).

	\begin{figure}[!t]
		\begin{center}
			\includegraphics[width=0.95\columnwidth]{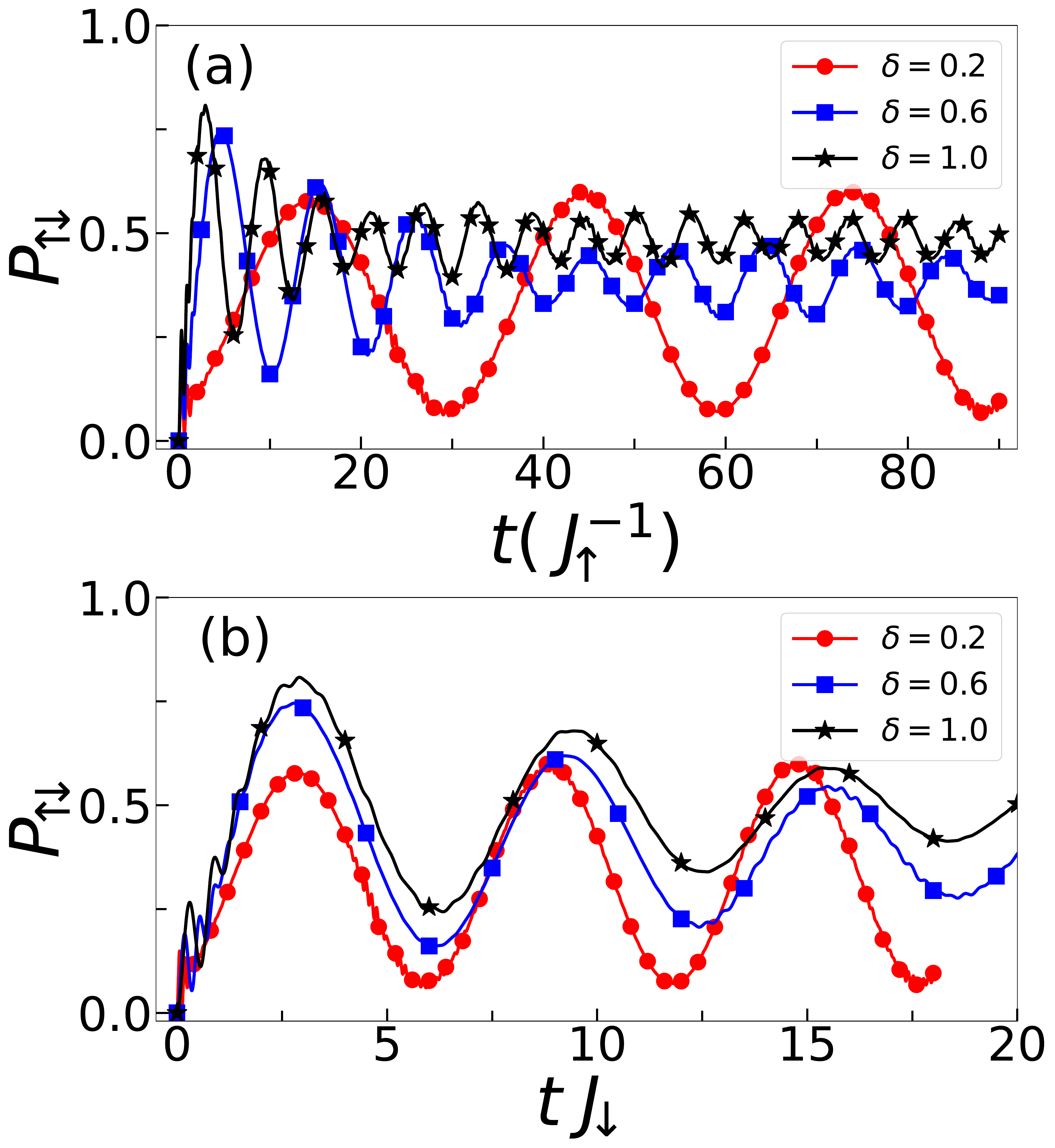}
		\end{center}
		\caption{Figure shows $P_{\ua\da}$ as a function of $t(J_\ua^{-1})$ (a), and $tJ_{\da}$ (b) for different $\delta$ with $U_{\ua\da} = U_{\ua} = U_\da = 10J_\ua$. }
		\label{fig:P_delta}
	\end{figure}
	
	We further analyze the situation at $U_{\ua\da} =10J_\ua$ by comparing the $P$'s for different values of $\delta$. In Fig.~\ref{fig:P_delta}(a) we show the variation of $P_{\ua\da}$ with time for $\delta = 0.2,~0.6$ and $1$. It can be seen that the period of oscillation increases with decreasing $\delta$. This phenomenon can be explained in the following way. When the two components form doublons, they move with an effective second-order hopping $\propto J_\ua J_\da/U_{\ua\da}$ as suggested by the perturbation theory~\cite{Kuklov2003,hsshh}. This leads to the different time-scales for different $\delta$ which is $\propto 1/J_\da$ (as other parameters are same for different $\delta$ cases)  in the QW. As a result we see different periods of oscillation in $P$'s with respect to time as shown in Fig.~\ref{fig:P_delta}(a). To confirm this we plot $P_{\ua\da}$ with $tJ_{\da}$ in Fig.~\ref{fig:P_delta}(b), which shows almost equal period of oscillation for different values of $\delta$. The slight deviation in the long time evolution for different $\delta$ can be attributed to the approximation in the perturbation theory. 
	
	\subsection{Nearest-neighbour pair}
\begin{figure}[!t]
		\includegraphics[width=1.0\columnwidth]{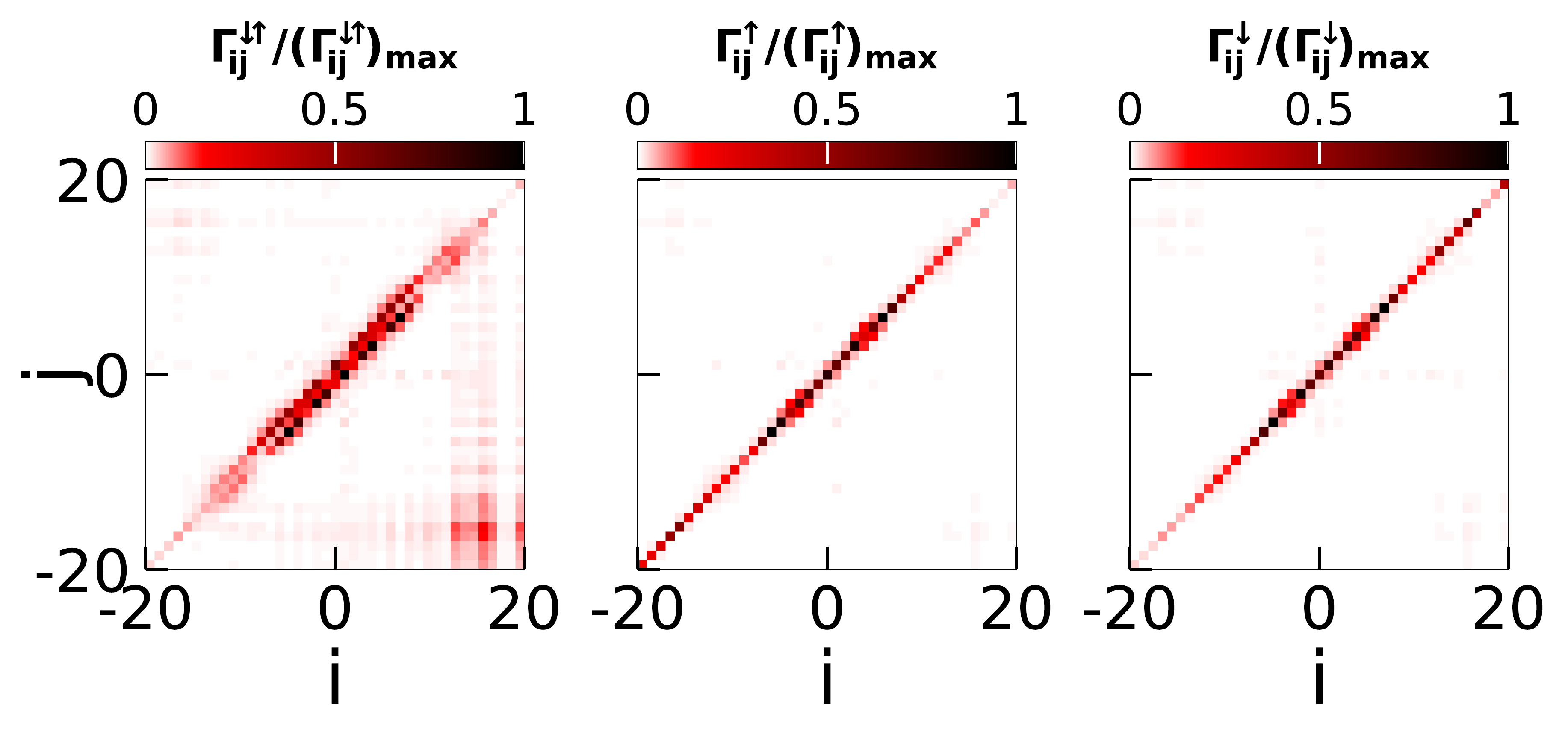}
		\caption{$\Gamma_{ij}^{\da\ua}$, $\Gamma_{ij}^{\ua}$ and $\Gamma_{ij}^{\da}$ are plotted at time $t=70J_\ua^{-1}$ for $U_{\ua\da}=U=10J_\ua$ and $\delta=1$. }
		\label{fig:corr_t2_1}
	\end{figure}

	As mentioned in the main text, we find the signatures of the nearest-neighbor bound pair at $U_{\ua\da} = 10J_\ua$. To confirm this we plot $\Gamma_{ij}^{\da\ua}$ at time $t=70J_\ua^{-1}$ in Fig.~\ref{fig:corr_t2_1} for $\delta = 1$. The finite upper and lower diagonal elements indicate the nearest-neighbor bound pair whereas the diagonal elements in $\Gamma_{ij}^{\ua}$ and $\Gamma_{ij}^{\da}$ ensure the formation of the $\ua\ua$ and $\da\da$ pairs. 
	\begin{figure}[!b]
		\begin{center}
			\includegraphics[width=1.0\columnwidth]{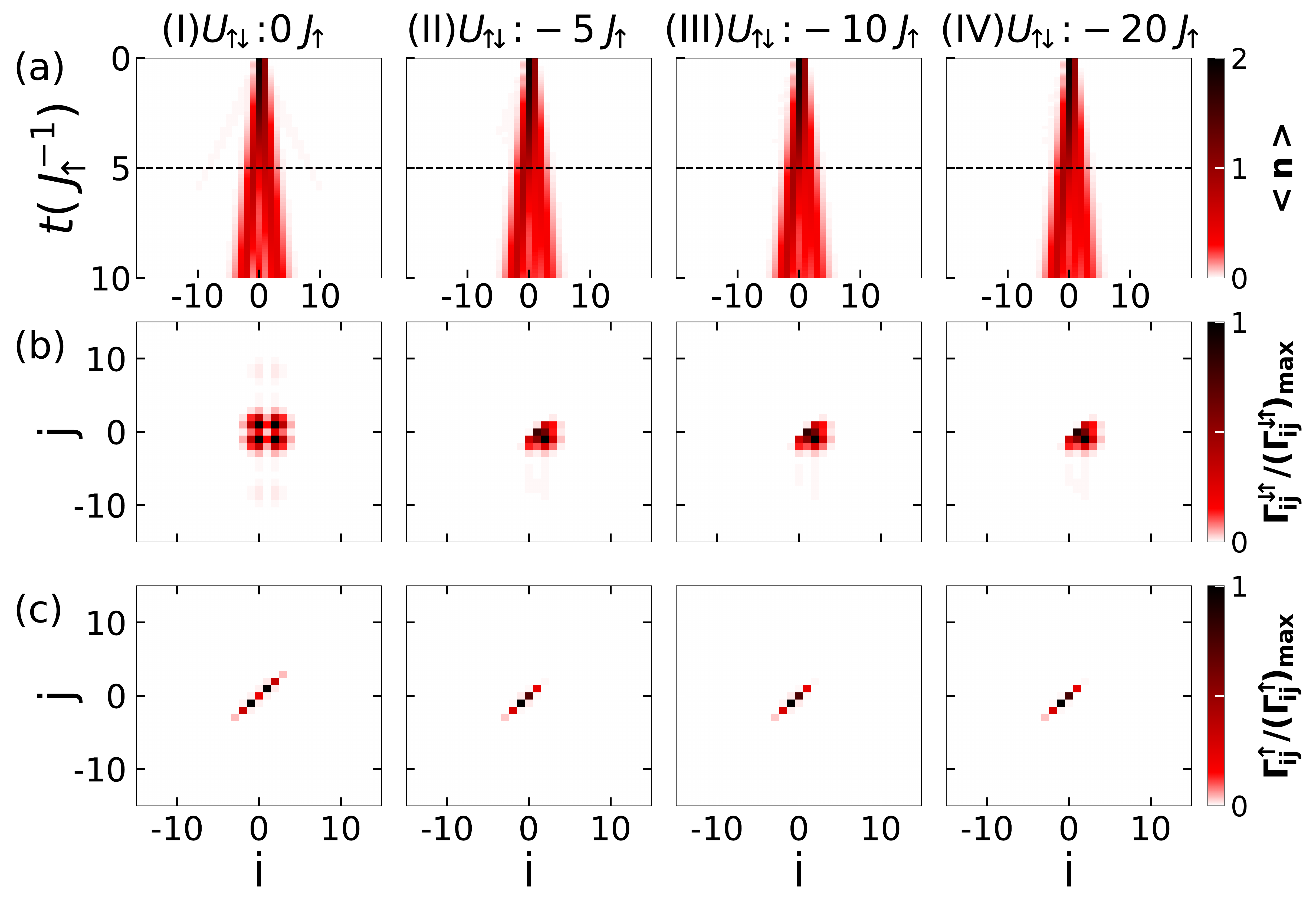}
		\end{center}
		\caption{Panel (a) shows the density evolution of the initial state $|\Psi_0\rangle = a_{0,\ua}^{\dagger 2}a_{1,\da}^{\dagger}|vac\rangle$ for $U_{\ua\da} = 0J_\ua$, $-5J_\ua$, $-10J_\ua$ and $-20J_\ua$. Here $U=10J_\ua$ and $\delta = 0.2$. Panel (b) and (c) are shown the correlation functions $\Gamma^{\da\ua}_{ij}$ and $\Gamma^{\ua}_{ij}$ respectively at time, $t=5J_\ua^{-1}$.}
		\label{fig:3den_cor_U_att}
	\end{figure}

		\begin{figure*}[!t]
		\begin{center}
			\includegraphics[width=2.0\columnwidth]{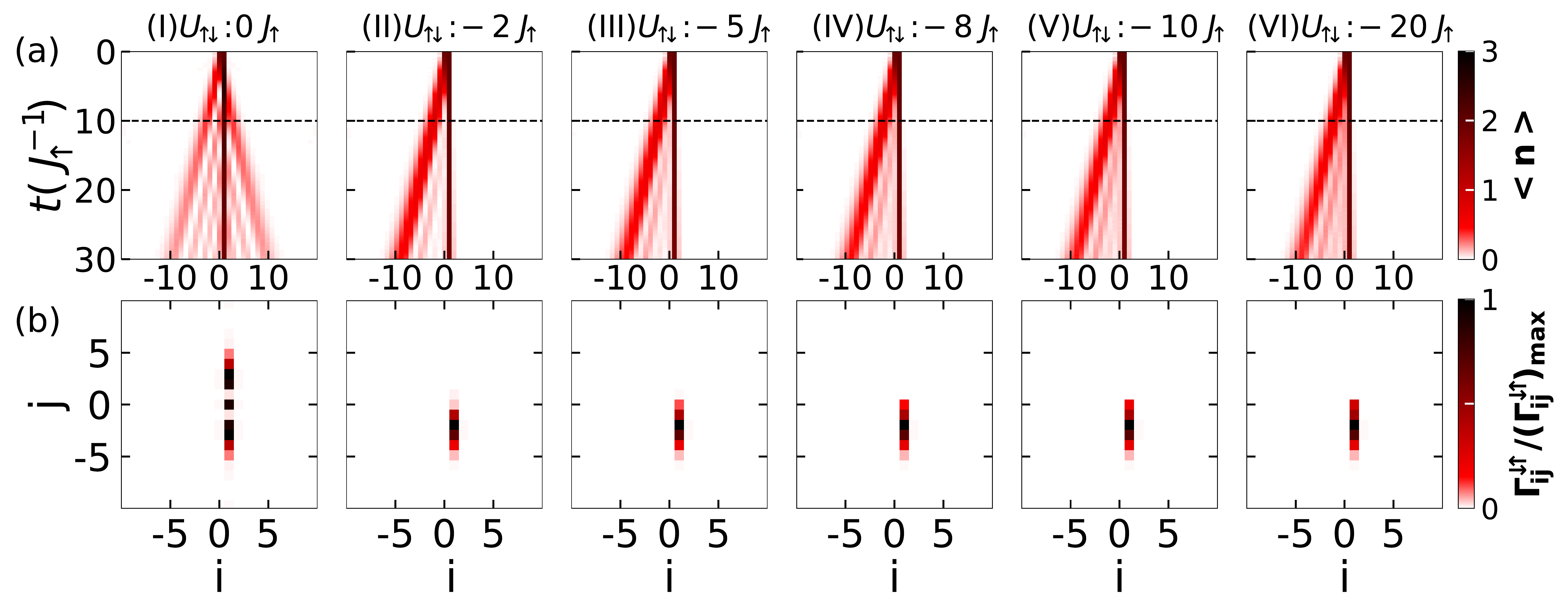}
		\end{center}
		\caption{Panel (a) shows the density evolution of the initial state $|\Psi_0\rangle = a_{0,\ua}^{\dagger 2} a_{1,\da}^{\dagger 2}|vac\rangle$ for $U_{\ua\da} = 0J_\ua, -2J_\ua, -5J_\ua, -8J_\ua$, $-10J_\ua$ and $-20J_\ua$. Here $U=10J_\ua$ and $\delta = 0.2$. Panel (b) shows the correlation functions $\Gamma^{\da\ua}_{ij}$ at time, $t=10J_\ua^{-1}$.}
		\label{fig:4den_cor_U_att}
	\end{figure*}		
	
\section{Repulsive U and attractive $U_{\ua\da}$}
The physics of bound state formation of non-local bosons discussed in the manuscript are are extremely sensitive to the nature of the interactions. The important condition that energetically favour these features in the dynamics demands all the associated interactions ($U_{\ua},~U_{\da}$ and $U_{\ua\da}$) repulsive or attractive in nature. If different signs are considered then the inter-component bound states do not form in the dynamics. As an example we consider repulsive intra component interactions ($U_{\ua}$ and $U_{\da}$) and attractive $U_{\ua\da}$ and study the QW using the initial states $|\Psi(0)\rangle_{\rm_{I}} = a_{0,\ua}^{\dagger 2} a_{1,\da}^{\dagger}|vac\rangle$ and $|\Psi(0)\rangle_{\rm_{II}} = a_{0,\ua}^{\dagger 2} a_{1,\da}^{\dagger 2}|vac\rangle$ in the following. 

\subsection{Two $\ua$ and one $\da$ particles}
The initial state we have considered here is $|\Psi(0)\rangle_{\rm_{I}} = a_{0,\ua}^{\dagger 2} a_{1,\da}^{\dagger}|vac\rangle$ and we plot the on-site density evolution $(\left\langle n_{i} \right\rangle )$, the density-density correlation ($\Gamma_{ij}^{\da\ua}$) and the two-particle correlation ($\Gamma_{ij}^{\ua}$) in Fig.~\ref{fig:3den_cor_U_att}(a), (b) and (c) respectively for $\delta=0.2$. Here we consider $U_{\ua} = U_{\da} = U = 10J_\ua$ and $U_{\ua\da}= 0J_\ua,~-5J_\ua,~ -10J_\ua,~-20J_\ua$. From Fig.~\ref{fig:3den_cor_U_att}, we can conclude that, unlike  the repulsive $U_{\ua\da}$ case, breaking of $\ua\ua$ pair and formation of $\ua\da$ pair is forbidden. The finite diagonal elements in the plot showing the two-particle correlation  matrics ($\Gamma_{ij}^{\ua}$) in Fig.~\ref{fig:3den_cor_U_att}(c) confirms the formation of only $\ua\ua$ pair for all values of $U_{\ua\da}\leq 0J_\ua$.

	\subsection{Two $\ua$ and two $\da$ particles}
	Now we consider the state $|\Psi(0)\rangle_{\rm_{II}} = a_{0,\ua}^{\dagger 2} a_{1,\da}^{\dagger 2}|vac\rangle$ as the initial state. We plot the on-site density evolution $(\left\langle n_{i} \right\rangle )$ and the density-density correlation ($\Gamma_{ij}^{\da\ua}$) in Fig.~\ref{fig:4den_cor_U_att}(a) and (b) respectively for $\delta=0.2$, $U_{\ua}=U_{\da}=U=10J_\ua$ and $U_{\ua\da}= 0J_\ua,~-2J_\ua, ~-5J_\ua,~ ~-8J_\ua,~ -10J_\ua,~-20J_\ua$. It can be clealry seen from Fig.~\ref{fig:4den_cor_U_att}(a) that the wavepacket of the $\ua\ua$ pair gets reflected from that of the strongly bound $\da\da$ pair for all finite values of $U_{\ua\da}$. Note that the slow spreading of the $\da\da$ is due to weak hopping strength. The reflection in this case can also be observed from the vanishing of the upper triangular matrix elements of $\Gamma_{ij}^{\ua\da}$ as depicted in Fig.~\ref{fig:4den_cor_U_att}(b). In this case we do not see any triplon or nearest neighbour pairing. 

	\section{Effective hopping of the triplon}

	In this section, we estimate the effective hopping strength of the triplon which is formed in the limit of large onsite interactions (see main text). To this end we employ the perturbative approach and find out the leading order approximation in the triplon hopping strength. We use a system consists of two site and three particles (two $\ua$ and one $\da$). The  Hamiltonian for this system can be written as,
	\begin{equation}
	H = H_{0}+H'
	\end{equation}
	where $H_{0}= U_{\ua\da}\sum_{i}n_{i,\ua}n_{i,\da}\nonumber + \sum_{i,\sigma} \frac{U_{\sigma}}{2} n_{i,\sigma}(n_{i,\sigma}-1)$ and $H'=-\sum_{\langle i,j\rangle,\sigma}J_\sigma(a_{i,\sigma}^{\dagger} a_{j,\sigma}+H.c.)$ which is considered as perturbation over $H_0$. 
	For the triplon hopping the relevant states are 
	$\left| 0 \;\;\;\; \da\ua\ua \right\rangle$ and $\left| \da\ua\ua \;\;\;\; 0 \right\rangle$. With these states and the above Hamiltonian the first order correction yields 
	\begin{equation}
	J_{eff}= -\frac{J_{\ua}^2J_{\da}}{(U_{\ua} + U_{\ua\da})^2} -\frac{J_{\ua}^2J_{\da}}{U_{\ua\da}(U_{\ua} + U_{\ua\da})} .
	\label{Eq:27}
	\end{equation}
Inserting the condition for the formation of triplon in our case which is $U_{\ua}=U_{\da} = U$ and $U_{\ua\da} = \frac{U}{2}$, we get
	\begin{equation}
	J_{eff} = -\frac{4J_{\ua}^2 J_{\da}}{9U_{\ua\da}^2}
	\label{Eq:28}
	\end{equation}
	
	So, the effective hopping of the triplon $J_{eff} \propto \frac{J_{\ua}^2 J_{\da}}{U_{\ua\da}^2}$.

	\bibliography{references}